# Future climate trends from a first-difference atmospheric carbon dioxide regression model involving emissions scenarios for business as usual and for peak fossil fuel

L.M.W Leggett and D.A. Ball

Global Risk Policy Group Pty Ltd

April 2014

**Abstract**

This paper investigates the implications of the future continuation of the demonstrated past (1960-2013) strong correlation between first-difference atmospheric $CO_2$ and global surface temperature. It does this, for the period from the present to 2050, for a comprehensive range of plausible future fossil fuel energy use scenarios. The results show that even for a business-as-usual (the mid-level IPCC) fossil fuel use estimate, global surface temperature will rise at a slower rate than for the recent period 1960-2000. Concerning peak fossil fuel, for the most common scenario the currently observed temperature plateau will in the near future turn into a decrease. The observed trend to date for temperature is compared with that for global climate disasters: these peaked in 2005 and are notably decreasing. The temperature and disaster results taken together are consistent with either a reduced business-as-usual fossil fuel combustion scenario into the future, or a peak fossil fuel scenario, but not with the standard business-as-usual scenario. If the future follows a peak fossil fuel pathway, a markedly decreasing trend in global surface temperature should become apparent over the next few years. If entertained, these results are evidence that the climate problem may require less future preventative action. If so, the same evidence is support for the case that the peak fossil fuel problem does require preventative action. This action is the same as it would have been for climate – the rapid transition to a predominantly renewable global energy system.

**Introduction**

Energy from fossil fuel is of fundamental importance to the functioning of society. It is axiomatic therefore that estimation of the range of possible future trends in fossil fuel production is essential for planning worldwide (OECD 1999; International Energy Agency 2013). First, the amount of future fossil fuel estimated to be available affects the mix of energy sources – fossil fuel or non-fossil fuel – estimated to be required for combustion for energy provision. If the future amount of fossil fuel is estimated as less than global demand, the gap between supply and demand must be closed. This can be achieved by demand reduction – societally difficult – or by supply increase, from non-fossil fuel sources. Second, negative externalities from emissions from fossil fuel burning - local (respiratory) and/or global (climate change) - will also vary depending on the future trend trajectory.

This last question, of the character of climate change expected from the atmospheric carbon dioxide contributed to by fossil fuel combustion, has taken on further complexity since Leggett and Ball (2014) showed that the *rate of change* of atmospheric carbon dioxide leads and is closely correlated to global surface temperature. The rate-of-change relationship means that temperature, and possibly other aspects of climate, are more sensitive to the change of atmospheric carbon dioxide than if the sensitivity were simply linear. For this reason, the question of future scenarios concerning atmospheric carbon dioxide deserves revisiting, because the same atmospheric-$CO_2$ future scenarios may have a more dynamic and immediate effect on climate variables than previously entertained.

This paper addresses these questions. It does so by the following means.

For the period up to 2050 the paper first from the literature outlines (a) a cross-section across the full range of proposed anthropogenic emissions scenarios; and (b) using these, by linear regression analysis derives indicative future atmospheric $CO_2$ levels. The paper then depicts future global surface temperature levels which have been published from previous modelling analyses; Finally, for the afore-mentioned range of indicative future atmospheric $CO_2$ levels, translates these into rates of change (in terms of the first derivative (first difference)). For the period 1850 to 2050, these past results to 2012 and future projections are compared with trends in climate observations, first for global surface temperature and second for global climate disasters.

**Methods**

To make it easier to visually assess the relationship between the key climate variables, most data are normalised using statistical Z scores (also known as standardised deviation scores) (expressed as “Relative level” in the figures). In a Z-scored data series, each data point is part of an overall data series that sums to a zero mean and variance of 1, enabling comparison of data having different native units.  See the individual figure legends for details on the series lengths. The period covered by the specific Z-score is shown on the figure as, for example, “Relative trend Z1960-2013”.

In the study, no attempt is made to translate results from Z-scores back to levels of temperature or numbers of climate disasters. This is because the aim is primarily to show trends and turning points, not, for example, to project specific values for global temperature.

The investigation is conducted using linear regression. Global atmospheric surface temperature and the annual number of global climate disasters are the dependent variables. For these two variables, we tested the relationship between (1) the level of atmospheric $CO_2$ and (2) the change in the level of atmospheric $CO_2$. We express these $CO_2$-related variables in terms of the first finite difference, which is a convenient approximation to the first derivative (Hazewinkel, 2013). It is noted that there is a considerable background to the use of the rate of change of atmospheric $CO_2$ – including in first difference form – in climate studies - for example, Kaufmann et al. (2006).

When  change in a data series is expressed as the result of subtracting the value for the previous year from that of the current year, the resulting series is termed a first differences series. It is noted that a first differences series differs from one involving percentage change in that the first differences series preserves the relative scale of the change.

Variability is explored using both intra-annual (monthly) data and interannual (yearly) data. The period covered in the figures is shorter than that used in the data preparation because of the loss of some data points due to calculations of differences and of moving averages.

The quantification of the degree of relationship between different plots was carried out using regression analysis to derive either the correlation coefficient (R) or the coefficient of determination ($R^2$) for each relationship. Student's t-tests were used to determine the statistical significance of the correlations.

Annual data is presented unsmoothed. For monthly data, smoothing methods are used to the degree needed to produce similar amounts of smoothing for each data series in any given comparison. Notably, to achieve this outcome, series resulting from higher levels of differences require more smoothing. Smoothing is carried out initially by means of a 13-month moving average – this also minimises any remaining seasonal effects. If further smoothing is required, then this is achieved by taking a second moving average of the initial moving average (to produce a double moving average). This is performed by means of a further 13 month moving average, to produce a 13 x 13 moving average.

**Data sources**

The following are the data series used in this analysis and their sources.

We used the Hadley Centre–Climate Research Unit combined land SAT and SST (HadCRUT) version 4.2.0.0 tropics (30S-30N) average http://www.metoffice.gov.uk/hadobs/hadcrut4/data/download.html. This series is used because (Leggett and Ball 2014) it shows the highest correlation with first-difference atmospheric $CO_2$. It is noted (Leggett and Ball 2014) that HADCRUT4 tropics is closely correlated with HADCRUT4 global surface temperature.

Atmospheric $CO_2$ data is from the U.S. Department of Commerce National Oceanic & Atmospheric Administration Earth System Research Laboratory Global Monitoring Division Mauna Loa, Hawaii monthly $CO_2$ series (annual seasonal cycle removed) ftp://ftp.cmdl.noaa.gov/ccg/CO2/trends/CO2_mm_mlo.txt.

Disaster information is from the WHO Collaborating Centre for Research on the Epidemiology of Disasters (CRED) Emergency Events Database EM-DAT (EM-DAT 2013). In the EM-DAT database, climate disaster types are sorted into four main categories: extreme temperature, flood, storm and wildfire. The extreme temperature category is further is made up of two main sub-types – heat wave and cold wave. In this study the heat wave sub-type is used. The number of events per year for each of the preceding categories is added to produce an annual climate disaster time series.

Unless otherwise stated, as long time series as reasonably practicable are used in this analysis. At its maximum, this period is 1850-2050. This perspective can provide the fullest possible indication of (i) the existence of common patterns and (ii) the relative scale of changes and their relative long-run frequency - including their relative unprecedentedness or otherwise.

Further details of data sources are given in Table 1.

Table 1. Data sources

| Data series | Content | Source | Internet location |
|---|---|---|---|
| Atmos. $CO_2$ (RCP4.5 scenario) | Atmos. $CO_2$ RCP4.5 scenario | | RCP DATABASE: http://www.iiasa.ac.at/web-apps/tnt/RcpDb |
| Lean and Rind temperature model | | Lean, J. L. & Rind D. H. How will Earth's surface temperature change in future decades? Geophys. Res. Lett. 36 L15708 (2009). | |
| BP world energy use projection | | Christof Ruhl: 'BP Energy outlook 2030' Statistical Review of World Energy; http://www.bp.com | |
| Proj temp mip5_global tas_Amon_modmean_rcp45 | IPCC RCP4.5 temperature projection | | http://climexp.knmi.nl/data/tsicmip5_tas_Amon_modmean_rcp45_0-360E_-90-90N_n_+++.txt |
| Anthro. $CO_2$ (Peak fossil fuel scenario) | Z1860-2011 Anthro. $CO_2$ | BP 2013, Nel and Cooper 2009, Laherrere 2012) | http://www.bp.com/statisticalreview, Nel and Cooper 2009, Laherrere 2012 |
| IPCC A2 temperature projection | IPCC TAR SRES A2 temperature projection | WDCC - World Data Center for Climate Hamburg | http://cera-www.dkrz.de/WDCC/ui/Compact.jsp?acronym=HADCM3_SRES_A2 |
| Total climate forcing incl. volc. (RCP4.5 scenario) | Z1860-2011 TOTAL_INCLVOLCANIC_RF | | RCP DATABASE: http://www.iiasa.ac.at/web-apps/tnt/RcpDb |
| IPCC 20C3M temperature dataset | Temperature dataset between 1765 and 2005 | Timespan: For convenience, RCP3PD, RCP45, RCP6 and RCP85 datasets, as well as two supplementary extension files include datasets between 1765 and 2005, identical to the provided 20c3m dataset. | |
| Drought | | EM-DAT: The OFDA/CRED International Disaster Database – www.emdat.net – Université catholique de Louvain – Brussels – Belgium (EM-DAT 2013) | http://www.emdat.be/disaster-list |
| Cold wave | | | |
| Heat wave | | | |
| Wildfire | | | |

## Results

In what follows, for clarity: (i) results are cumulated iteratively, one result at a time; and (ii) in the figures each new result added to the pre-existing group of curves is depicted in red.

Results are grouped under three subheadings: global surface temperature; global climate disasters; and future scenarios for global fossil fuel consumption.

### *Global surface temperature*

Figure 1 shows data for the mid-range (van Vuuren et al. 2011) IPCC representative concentration pathway RCP4.5 scenario. The figure shows: modelled atmospheric carbon dioxide; and the projected global surface temperature for the multi-model means for models run (a) during 2005 and 2006 (http://cmip-pcmdi.llnl.gov/cmip3_overview.html?submenuheader=1) for the 2007 Fourth Assessment Report (AR4) of the Intergovernmental Panel on Climate Change (IPCC) (phase 3 of the Coupled Model Intercomparison Project (CMIP3) and (b) up to generally 2012 for the Fifth Assessment Report (AR5,2013) (CMIP5). (The Representative Concentration Pathways (RCPs) are a set of four new pathways developed for the climate modelling community as a basis for long-term and near-term modelling experiments
The four RCPs together span the range of radiative forcing values calculated to year 2100 found in the open literature.)

The group of curves in Figure 1 shows the very close agreement between the model for $CO_2$ and both the two models CIMP3 (earlier) and CIMP5 (more recent). The graph shows the dominance of the temperature projections by the linear level of $CO_2$. Given this close similarity, for simplicity in the following figures the linear RCP4.5 $CO_2$ projection is used to indicate the broad trajectory of the IPCC mid-range bundle of curves. The figure shows that both the earlier model averages run in the 2006-7 period and the later from 2012 are closely similar to each other – the later model average being slightly higher to 2030 -  and that both the overall models are throughout dominated by the level of atmospheric $CO_2$.

Figure 1. The relative trend (Z-scores) in historic and simulated time series of atmospheric carbon dioxide and the anomalies in annual global-mean surface temperature for the mid-range (van Vuuren et al. 2011) IPCC representative concentration pathway RCP4.5 scenario. All simulations use historical data up to and including 2005 and use RCP4.5 after 2005.  The figure shows: modelled atmospheric carbon dioxide (blue curve); and the projected global surface temperature for the multi-model means for models run (a) during 2005 and 2006 (http://cmip-pcmdi.llnl.gov/cmip3_overview.html?submenuheader=1) for the 2007 Fourth Assessment Report (AR4) of the Intergovernmental Panel on Climate Change (IPCC) (phase 3 of the Coupled Model Intercomparison Project (CMIP3) (purple curve) and  (b) up to generally 2012 for the Fifth Assessment Report (AR5,2013) (CMIP5)(yellow curve).

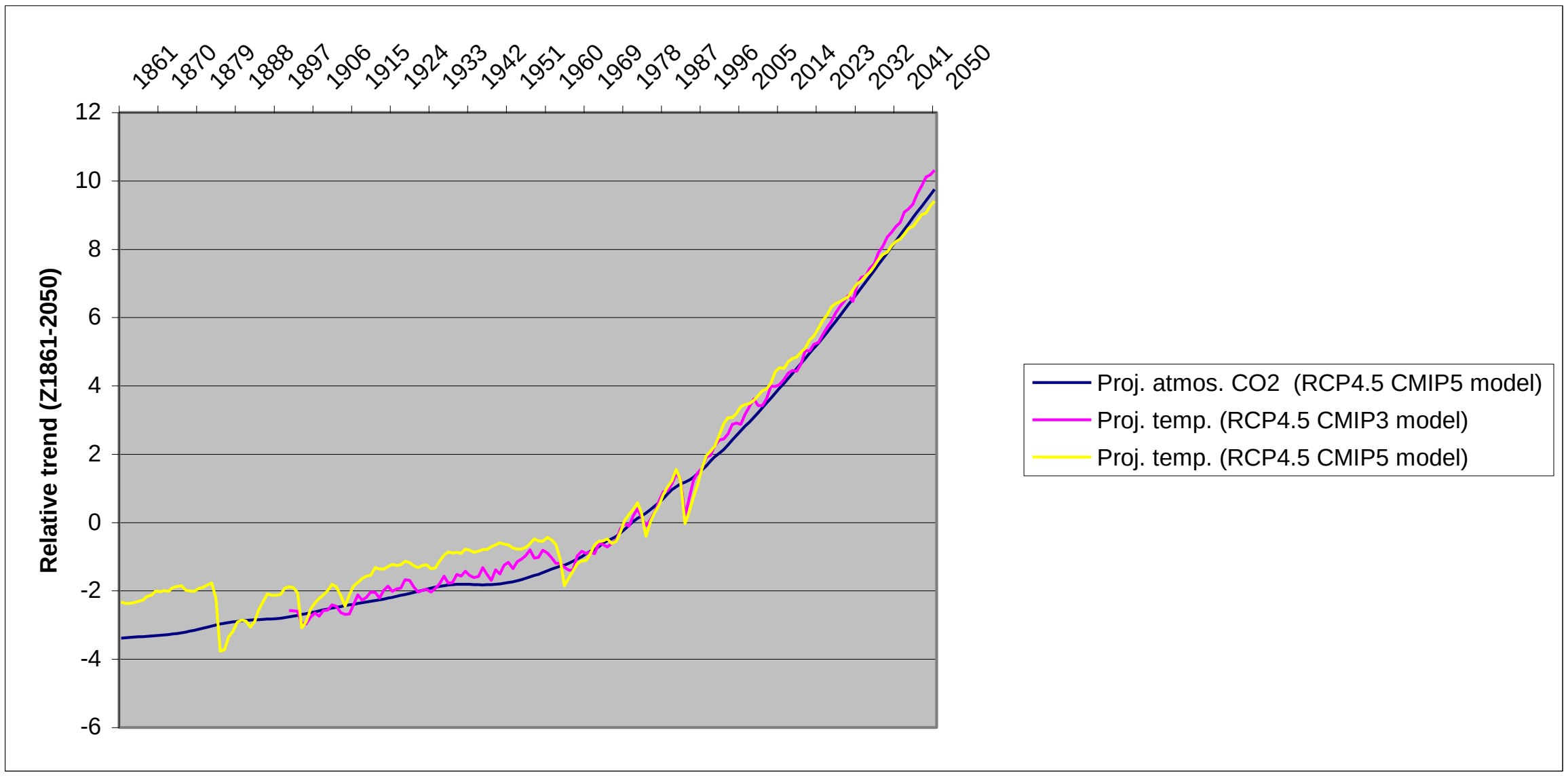

Figure 2 utilises the RCP4.5-scenario-modelled atmospheric carbon dioxide from Figure 1 to represent both itself and, given their demonstrated close similarity shown in Figure 1, the two RCP4.5 temperature series. Added to this in Figure 2 are the observed global surface temperature and the observed seasonally adjusted level of atmospheric carbon dioxide as measured at the Mauna Loa atmospheric station (NOAA, 2013).

Figure 2 illustrates the start point of this study: the familiar and much discussed topic that over the last 10 years or so the temperature trend has started to statistically significantly diverge from the average projection of current climate models (Fyfe et al. 2013). Figure 2 shows that the trend in observed atmospheric carbon dioxide to 2012 does not throw light on this point: it closely follows – in fact is slightly higher than - the IPCC temperature trend.

Figure 2: RCP4.5-scenario-modelled atmospheric carbon dioxide from Figure 1 (blue curve) standing for itself and, given their demonstrated close similarity, the two RCP4.5 temperature series shown in Figure 1; observed seasonally adjusted level of atmospheric carbon dioxide as measured at the Mauna Loa atmospheric station (light blue curve); and observed global surface temperature (Hadcrut4; ref) (red curve) for the period 1850 to 2013.

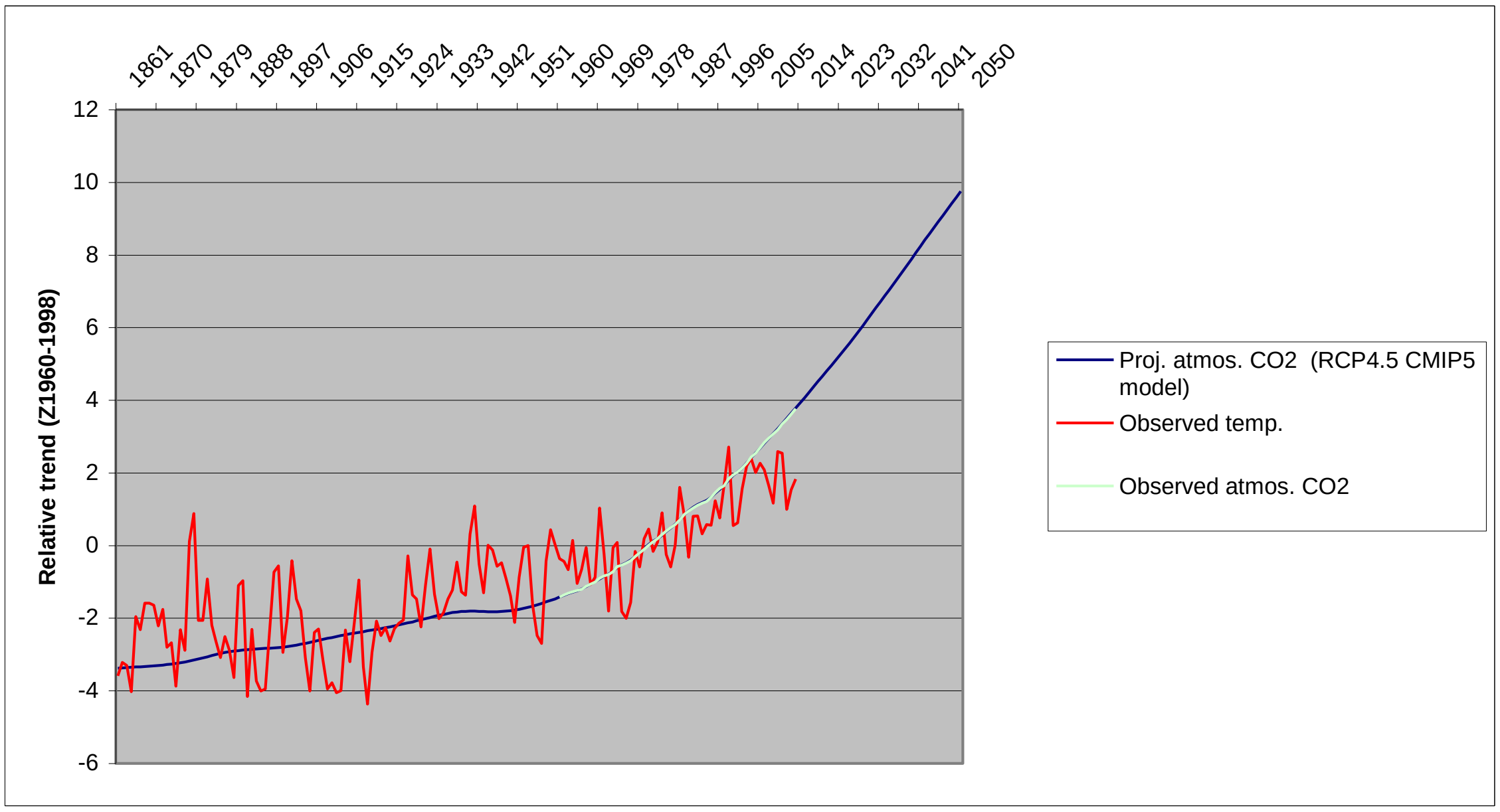

A range of models other than the IPCC projections exist. An illustrative example is the model of Lean and Rind (2009). The trend projected for this model is added to the suite of trends from the previous figures in Figure 3. This trend is lower than the IPCC mid-range projections, but higher than the observed temperature trend in recent years.

Figure 3. Data as for Figure 2 except that the curve for observed global surface temperature (Hadcrut4; ref) 1850 to 2013 is now depicted in lilac; and the global surface trend projected to 2030 from the regression model of Lean and Rind (2009) is added (red curve).

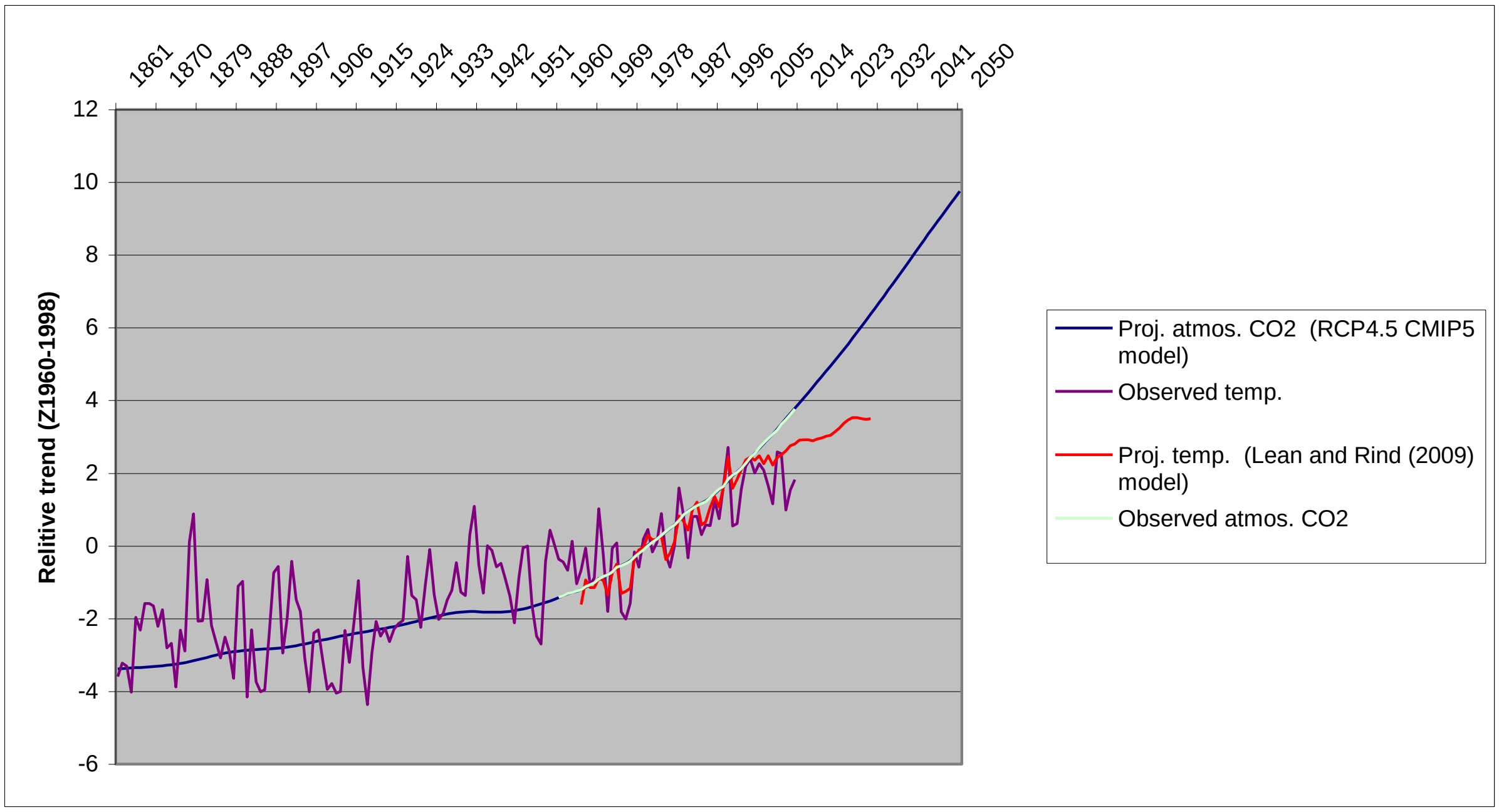

Figure 4 shows however (Leggett and Ball 2014), that the annual change in atmospheric carbon dioxide -- expressed in terms of first differences -- *does* show a levelling off similar to that of temperature.

Figure 4. Data as for Figure 3 except that the global surface trend projected to 2030 from the regression model of Lean and Rind (2009) is depicted in light blue; and the smoothed (13 month moving average) first difference of observed atmospheric $CO_2$ is added (red curve).

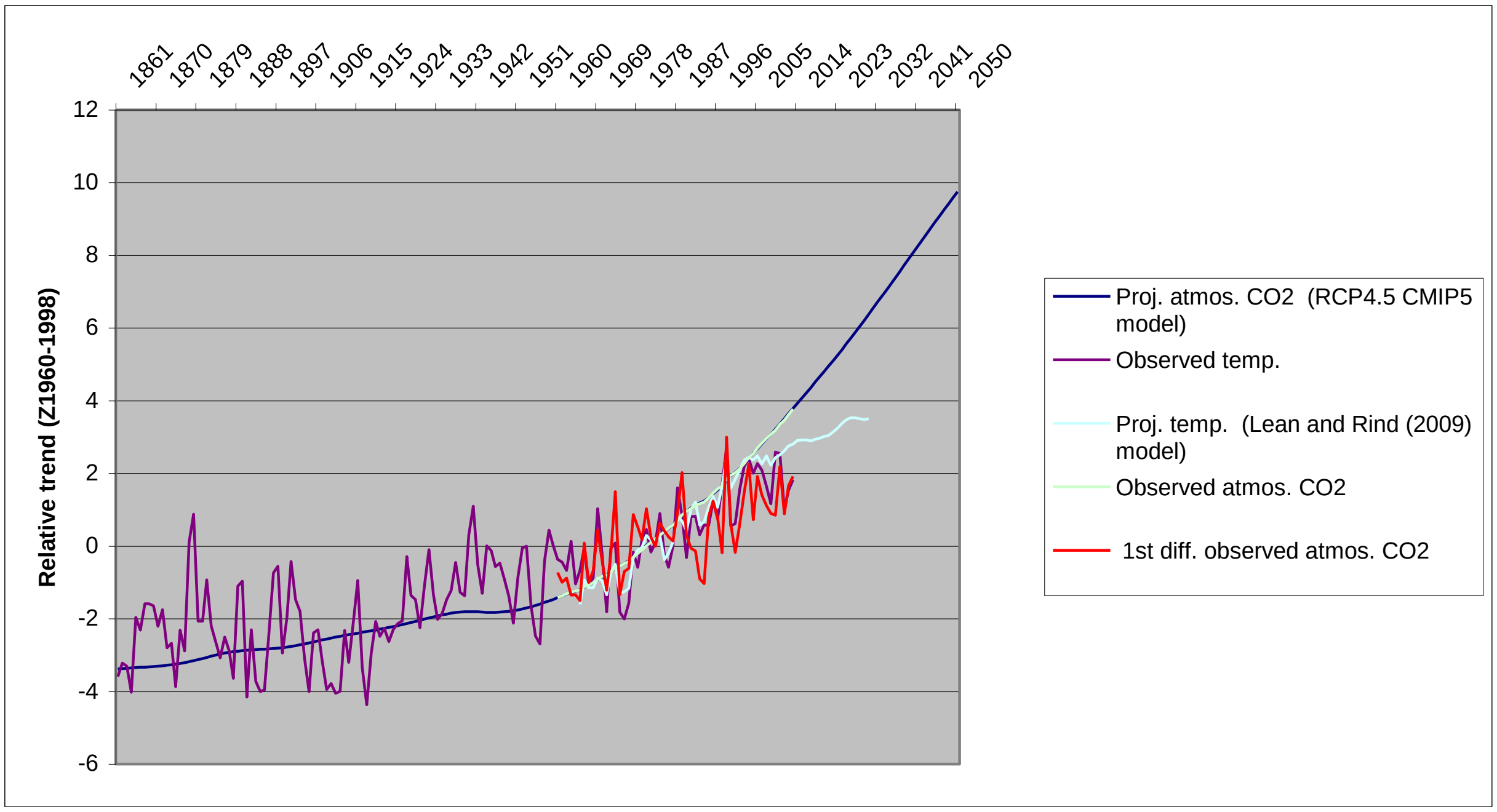

The relationship in monthly terms between first-difference atmospheric carbon dioxide and global surface temperature is shown from 1958 to 2013 in Figure 5.

Figure 5. 1958 to 2013 monthly: First-difference atmospheric carbon dioxide (smoothed by a 13 month moving average) (dark blue curve) and global surface temperature (purple curve)

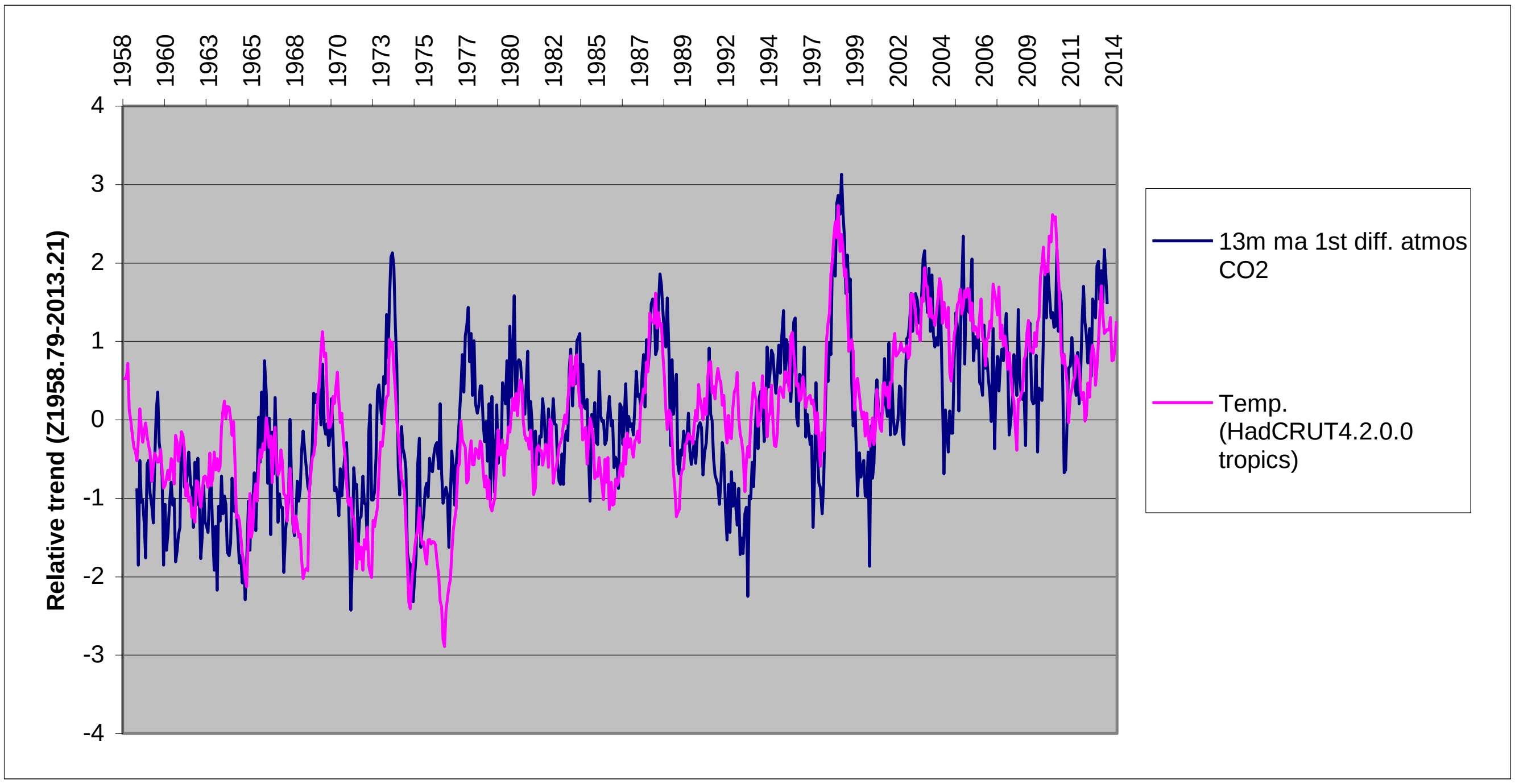

As shown in Leggett and Ball (2014), but with data now updated to early 2013, the enlarged figure shows clearly the very close similarity between the first-difference carbon dioxide and the temperature signatures. Visual inspection leaves very much the impression that the one curve may depart from the other from time to time but that if this happens the curves soon come back into synchronisation again. This is true over the entire period of the data involving some 600 data points. Using an accepted convention (Comer and Gould, 2011) the degree of correlation is considered strong ($R = 0.70$), and the statistical significance of the correlation is extremely high ($P=7.02E\text{-}98$). Correcting for autocorrelation with the Cochrane-Orcutt procedure leads after four iterations to a Durbin-Watson statistic of 2.08 (showing little or no remaining autocorrelation) and a lower statistical significance, but one which is still high, of 0.00011.

### *Global climate disasters*

Next, what is the situation for the trend concerning climate outcome categories other than temperature? One such group is climate disasters, for which the standard categories are wildfire, heat wave, drought, flood and storm (EM-DAT, 2013). In considering the trend for climate disasters it is recognised (for example, Guha-Sapir and Below, 2002) that for earlier decades due to poorer reporting there may be greater amounts of missing data. To minimise issues concerning possible missing data in earlier decades, based on an empirical assessment (see Supplementary Information), data used in the following section is (i) limited to OECD countries and (ii) commences in 1960. That said, the series for both OECD countries and the rest of countries are similar, and in particular (Supplementary Figure 7) each shows peaking in climate disasters between 2000 and the present.

Figure 6 shows the trend in counts per year for OECD countries from 1960 to 2013 for each of wildfire, heat wave, drought, flood and storm (EM-DAT 2013).

Figure 6. OECD countries: number of disasters per year from 1960 to 2013 for each of wildfire, heat wave, drought, flood and storm (EM-DAT 2013).

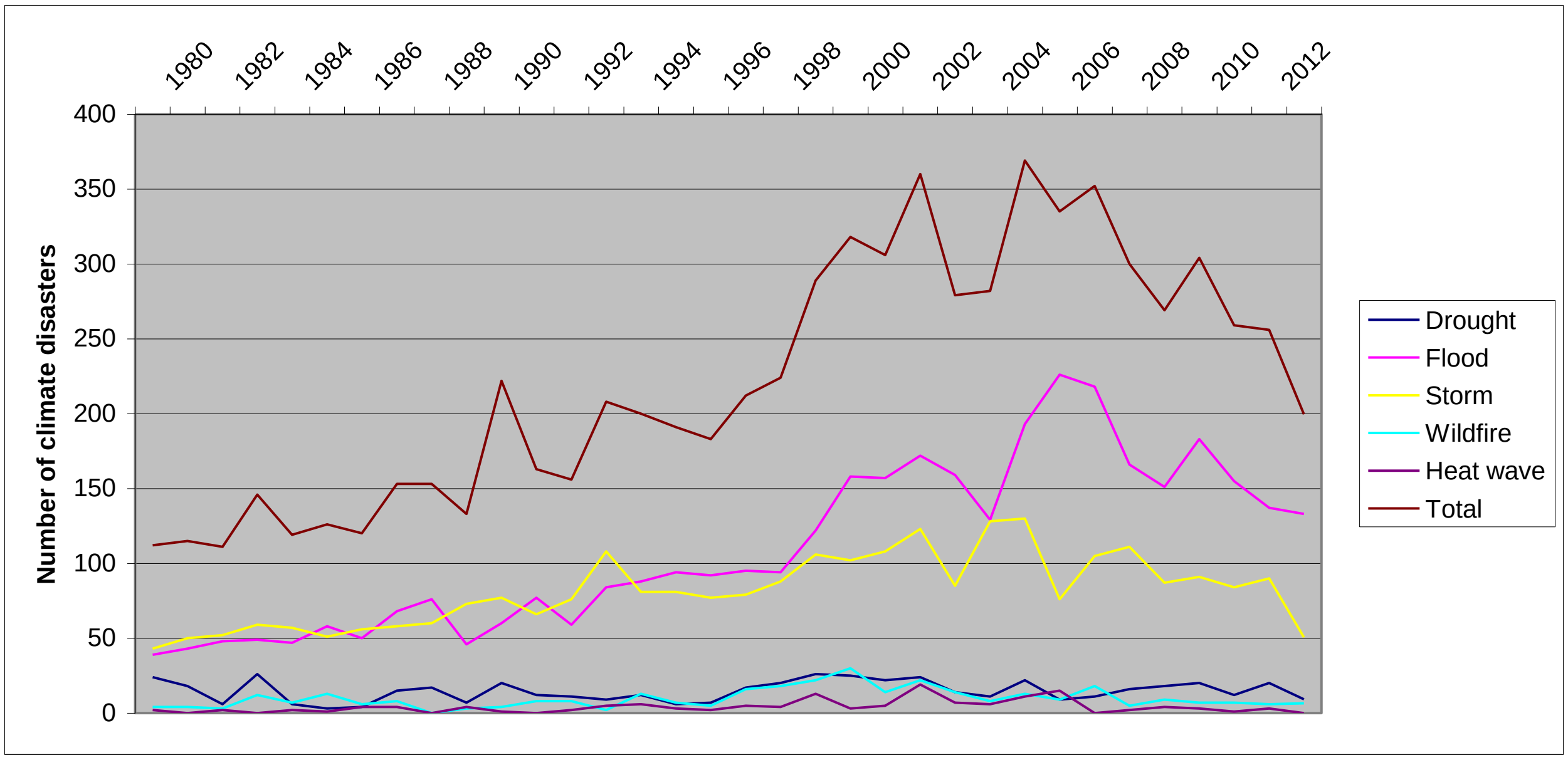

The figure shows that, although the number of events per year and timing of peaks vary, all disaster types without exception show a peak and then a decrease.

The counts for individual OECD climate disaster types are summed into a total category termed OECD climate disasters. In Figure 7 this trend is superimposed over those depicted in Figure 4.

Figure 7. Data as for Figure 4 except that the smoothed (13-month moving average) first difference of observed atmospheric $CO_2$ is depicted in mid blue; and observed total OECD climate disasters are added (red curve).

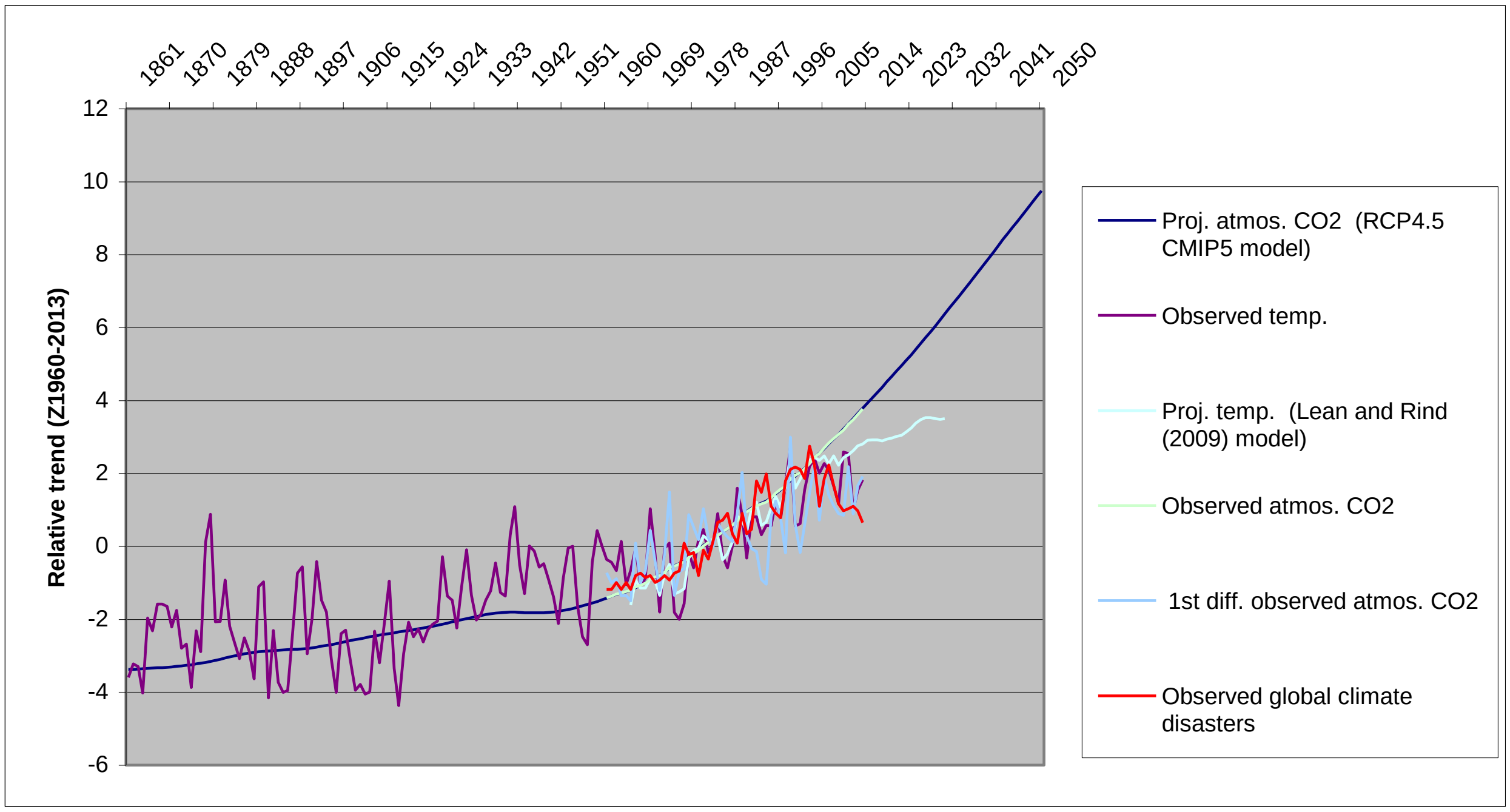

Figure 7 shows that the OECD climate disaster trend for 1960 to 2013 shows a clear peak (around 2005) followed by a steady decrease since then. Such a peak and decline is unprecedented over the half-century period depicted. The slope of this decreasing trend since 2005 is seen to depart even further from the linear $CO_2$ trend and that proposed for temperature by Lean and Rind (2009) than the previously shown global surface temperature and atmospheric $CO_2$ curves.

With the trends depicted above visually in terms of descriptive statistics, statistical tests are now carried out on the strength and statistical significance of the relationships between the trends.

Two characterisations are made: each of the two outcomes of temperature and climate disasters is assessed against both linear and 1st difference atmospheric $CO_2$.

What are the equivalent statistical results for the range of models depicted in the earlier figures? To avoid use of non-linear models, linear regressions are conducted for the year from peak first-difference atmospheric $CO_2$ and temperature – 1998 – for the succeeding years up to the latest year available – 2013. Annual data is used. Figure 8 illustrates the data for the period.

Figure 8. Depiction of annual data from 1998 to 2013 used for linear regression analysis to assess statistical significance of differences between observed and modelled climate trends

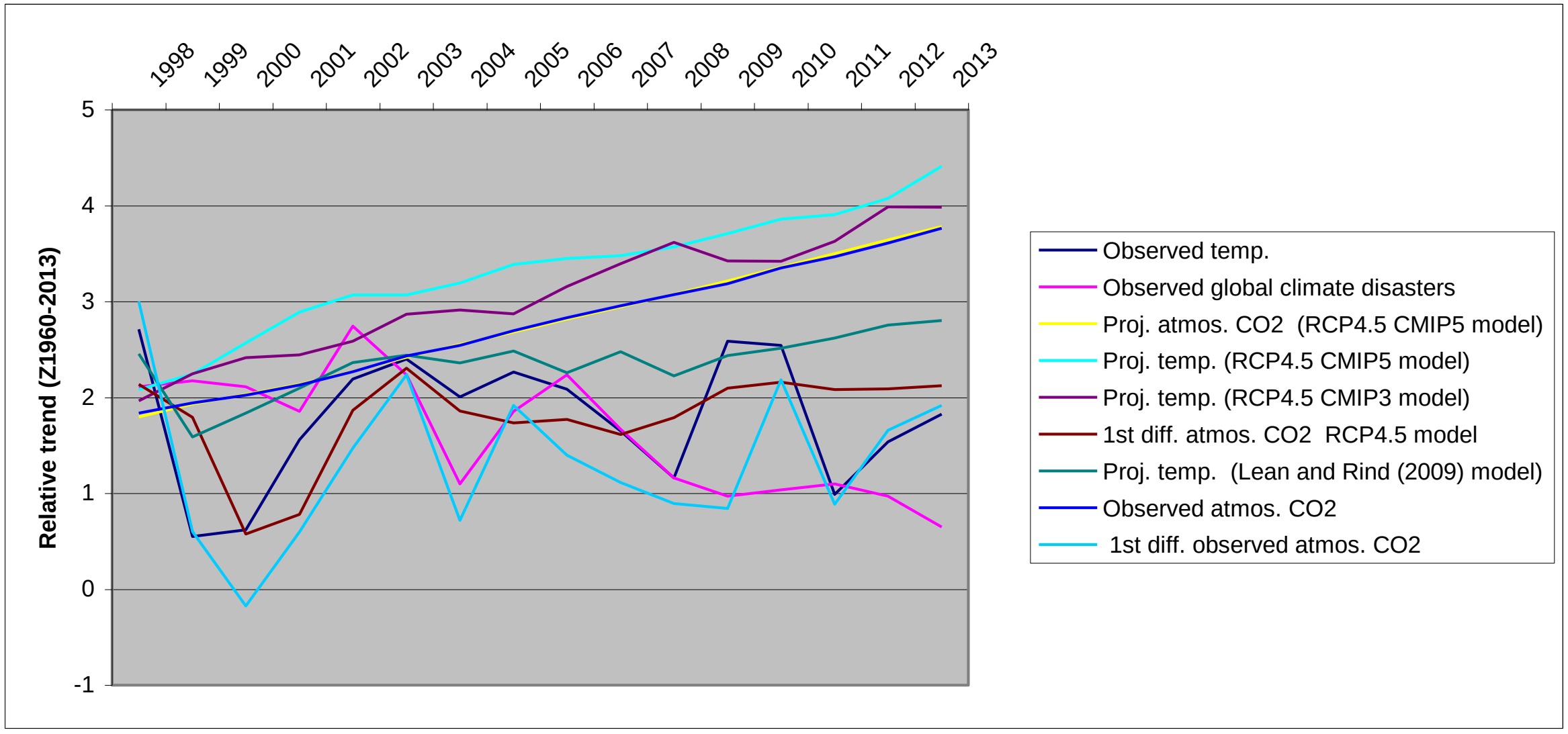

Correlation coefficients for and the statistical significance of the relationships between the trends in Figure 8 are displayed in Tables 2 and 3.

Table 2. Correlation coefficients for and the statistical significance of the relationships between the observed and modelled temperature trend. Green: substantial correlation and/or statistically significant; orange: not statistically significant.

| **Predicted temperature from**: | **Observed temperature** | |
|---|---|---|
| | R | P |
| Projected temp CMIP3 multi-model mean | 0.000 | 0.999842 |
| Linear atmos. $CO_2$ RCP4.5 model | 0.074 | 0.78545 |
| Observed linear atmos. $CO_2$ | 0.083 | 0.758663 |
| Projected temp CIMP5 multi-model mean | 0.133 | 0.606927 |
| Temp model (Lean and Rind 2009) | 0.547 | 0.028387 |
| 1st diff. observed atmos. $CO_2$ | 0.734 | 0.001209 |

Table 2 shows that the largest correlation coefficient (R=0.734) between a $CO_2$ model and temperature is for the correlation between temperature and the first-difference atmospheric $CO_2$ model. This correlation is statistically significant (P= 0.0012).

Table 3 shows that the relationships between aggregate climate disasters and all but first-difference atmospheric $CO_2$ display strong correlations but have negative coefficients. Each of these relationships is highly statistically significant. As climate disasters are count data, results are from log-linear regression (Poisson distribution selected).

Table 3. Correlation coefficients for and the statistical significance of the relationships between temperature models and the observed and the observed trend in global climate disasters. Green: substantial correlation and/or statistically significant; orange: correlation of wrong sign or not statistically significant.

| **Independent variable** | | **Global climate disasters** | | |
|---|---|---|---|---|
| | X parameter | P (Poisson) | (Chi-square (LR) ) | Pseudo R-squared (Cox and Snell) |
| Projected temp. CMIP3 multi-model mean | -0.294 | < 0.0001 | 22.755 | 0.759 |
| Linear atmos. $CO_2$ RCP4.5 model | -0.305 | < 0.0001 | 24.815 | 0.793 |
| Observed linear atmos. $CO_2$ | -0.310 | < 0.0001 | 24.608 | 0.790 |
| Projected temp CIMP5 multi-model mean | -0.267 | < 0.0001 | 21.387 | 0.732 |
| Temp model (Lean and Rind 2009) | -0.394 | 0.001 | 11.599 | 0.503 |
| 1st diff. observed atmos. $CO_2$ | -0.003 | 0.952 | 0.004 | 0.000 |

Table 3 and Figure 8 show that there is little correlation between climate disasters and temperature and first-difference $CO_2$ for the period after the 1998 peak. For the full period for which climate disaster data is used however - 1960 to 2013 – a clear similarity in trend is seen (Figure 9). The correlation coefficient for this relationship is high (Pseudo $R^2$ (Cox and Snell) = 0.998), and the relationship is highly statistically significant ($P<0.0001$)).

Figure 9. Annual tropical surface temperature and global climate disasters 1960-2013

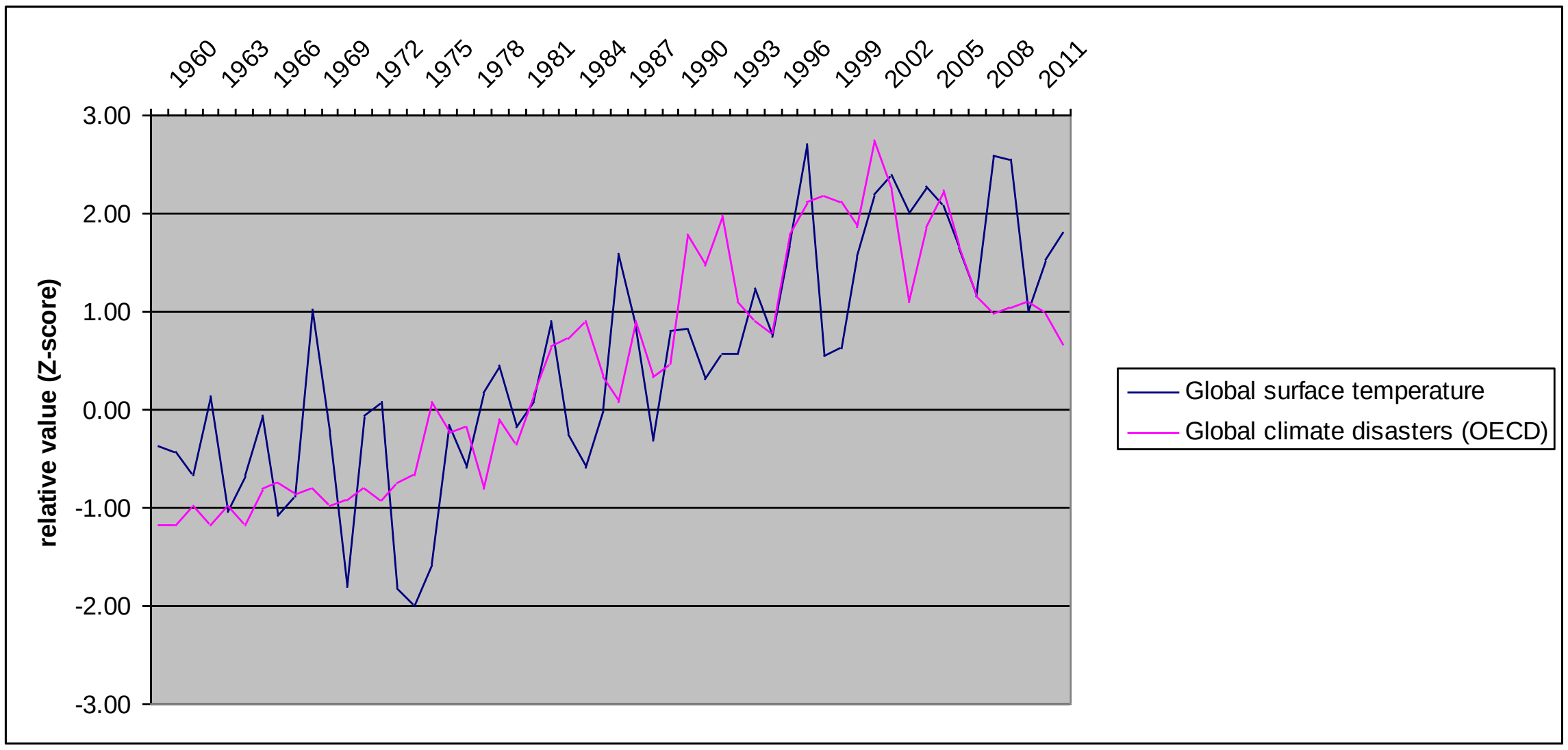

This study has shown, then, statistically significant evidence that, after 1998, two climate outcomes – global surface temperature and global climate disasters - on the one hand no longer increase with the increasing linear atmospheric $CO_2$ trend and on the other - for temperature - do follow the first-difference atmospheric $CO_2$ trend. For climate disasters the trend decreases statistically significantly as linear $CO_2$ increases. It is also shown that, as for temperature, it is likely that climate disasters are also following first-difference $CO_2$.

***Effect of future scenarios for global fossil fuel consumption***

With this evidence presented for links between climate outcomes and rate of change of $CO_2$, what might plausible future scenarios look like, first for the trend for atmospheric $CO_2$, and then for resulting climate outcomes under a first-difference $CO_2$ model?

The major modern source of atmospheric $CO_2$ is emissions from the combustion of fossil fuels. The RCP4.5 atmospheric $CO_2$ trend reflects the notion that fossil fuel production into the future will roughly follow demand and continue to rise (Meinshausen et al., 2011). Other business-as-usual models also exist. A key element of current modelling is the extent to which "unconventional” fossil fuels such as gas generated by hydraulic fracturing and oil from sources such as tar sands may change future expected global fossil fuel production. One business as usual study which explicitly incorporates unconventional fossil fuels is that of BP (2013). There has been a substantial and statistically significant linear correlation between anthropogenic $CO_2$ emissions and atmospheric $CO_2$ from 1959 to 2012 (Supplementary Information Figure 9). If the proposed BP (2013) output per year to 2030 translates by the relationship to date to the amount of carbon dioxide per year present in the atmosphere (for derivation see Supplementary Information, section 2), the following curve (red) in Figure 10 results.

It can be seen that this current (2013) estimate is - even allowing for unconventional fossil fuels - for lower emissions growth than expected in the RCP4.5 model. As well, a slightly but steadily decreasing growth rate is projected in the BP (2013) study.

Figure 10. Data as for Figure 6 except that total observed OECD climate disasters is depicted in pink and projected atmospheric $CO_2$ derived from projected anthropogenic $CO_2$ emissions (BP2013) is added (red curve).

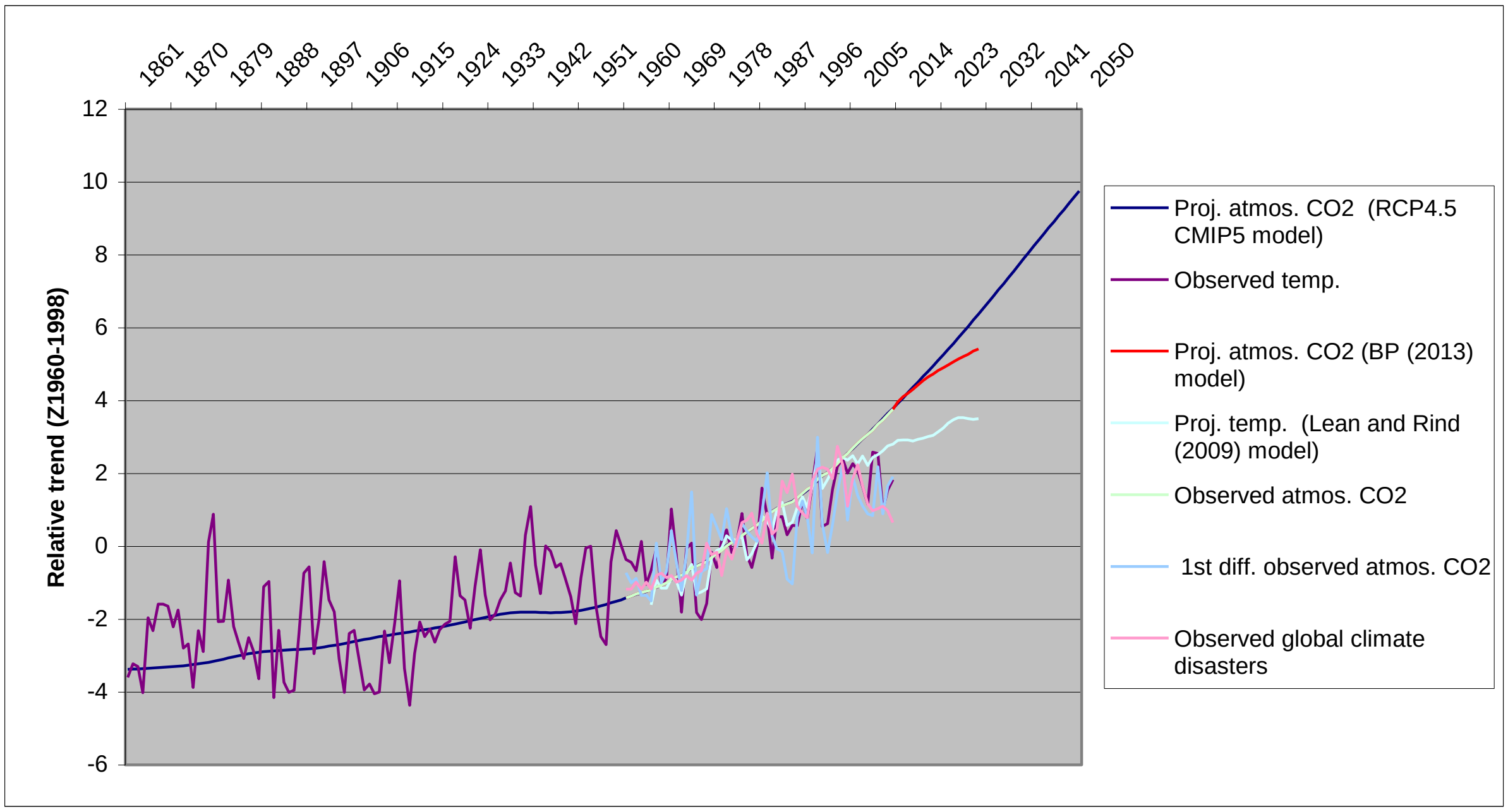

The alternate view to that of the IPCC and BP (2013) is that fossil fuel production will peak.

Figure 11 adds to the trends depicted in previous figures one (red curve) based on a peak fossil fuel scenario. This scenario is that of Nel and Van Zyl (2010) modified for the present study after Laherrere (2012) to allow for recent amended estimates of fossil fuel production from unconventional sources of the types assessed by BP (2013). Even with the addition of unconventional fuels, the Laherrere (2012) projection is that global fossil fuel is expected to peak in the mid-2020s and then to show a gradual decline. This peak fossil fuel trajectory is given in Figure 11.

It is noteworthy that the peak fossil fuel scenario shows a peak like that of the climate events, but it is some 20 years into the future (and see below).

Figure 11. Data as for Figure 9 except that projected atmospheric $CO_2$ derived from projected anthropogenic $CO_2$ emissions (BP 2013) is depicted in mid blue and projected atmospheric $CO_2$ (peak fossil fuel model) (see text) is added (red curve).

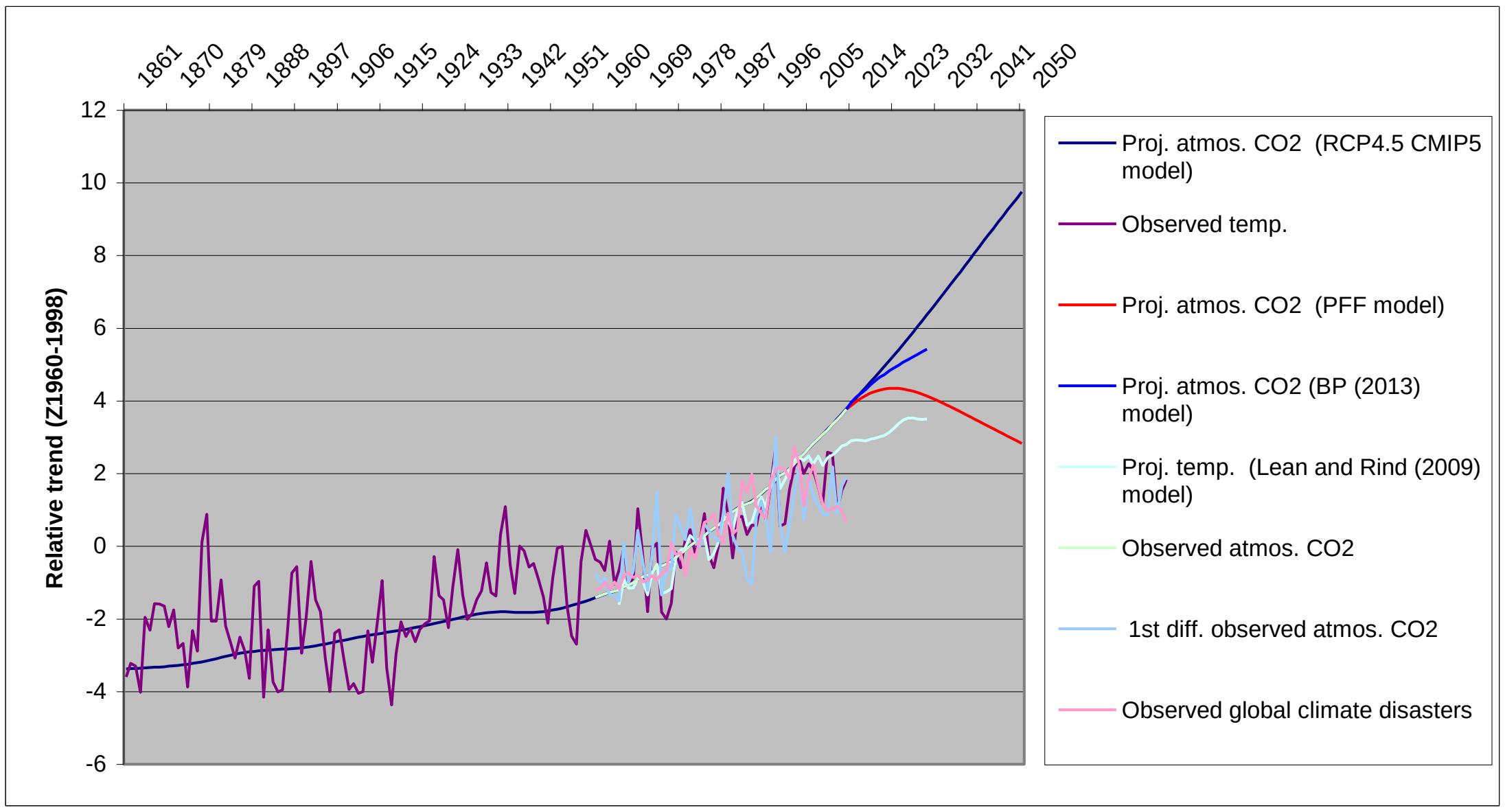

The following figures take the linear atmospheric $CO_2$ future level in Figure 8 and previous figures and express these as growth in first difference terms analogous to that shown for the actual data for the period 1958 to 2013 in Figures 8 and 9.

In Figure 12 the first difference of the IPCC atmospheric $CO_2$ estimated trajectory for the RCP 4.5 is given.

Figure 12. Data as for Figure 10 except that projected atmospheric $CO_2$ (peak fossil fuel model) (see text) is depicted in purple and the first-difference transformation of the RCP4.5 (CIMP5) model is added (red curve).

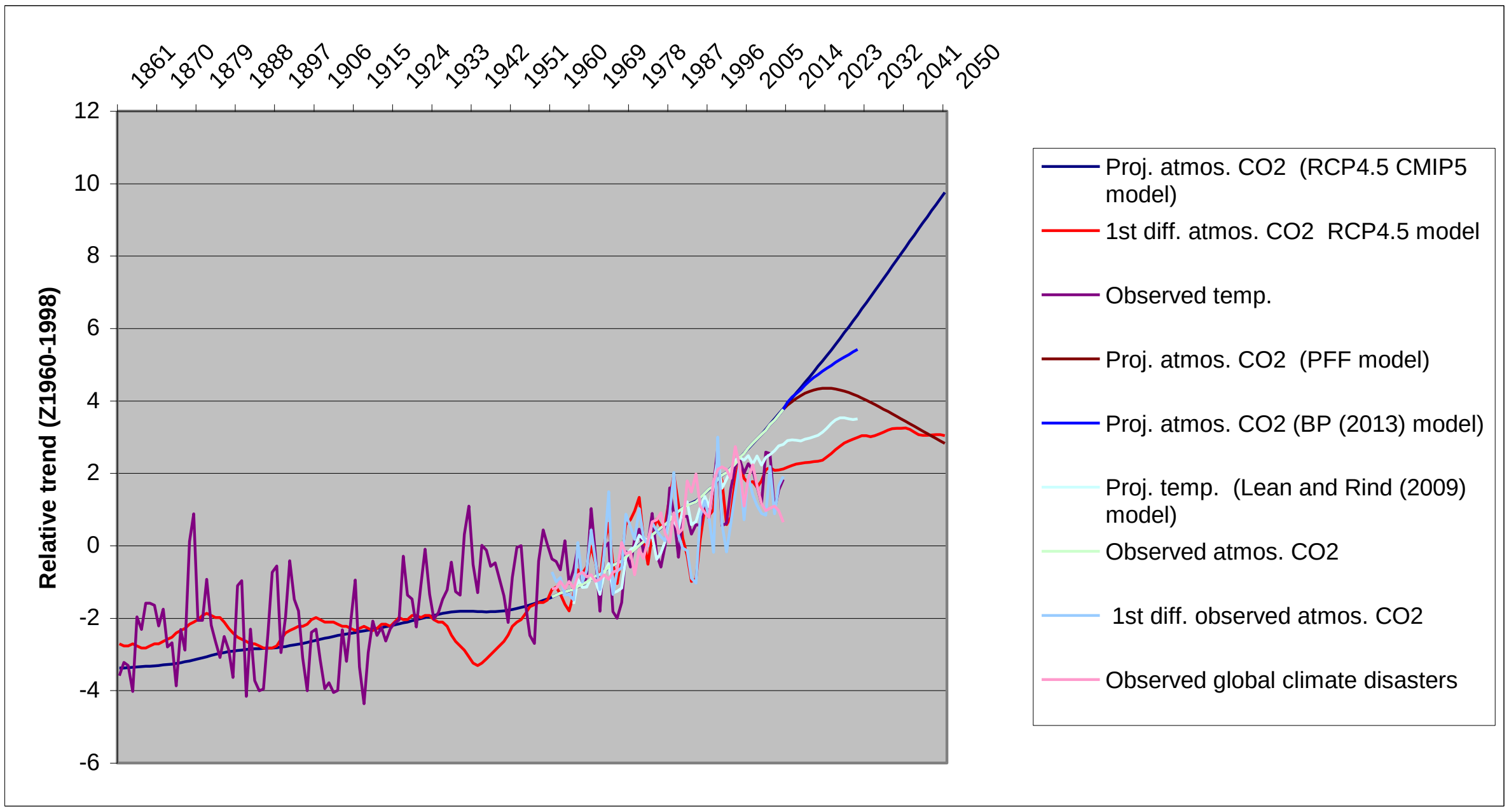

Figure 12 shows that the first difference RCP 4.5 trajectory reaches a maximum at about 2030 and then shows a slight decrease thereafter. It is noted that the RCP4.5 future trajectory is modelled as a smooth curve and thus its first difference shows less interannual variation than generated from the observed atmospheric $CO_2$ data introduced in figure 8. That said, this is not important for the purpose of this paper, which is to estimate broad trajectories into the future – up or down, more or less, not year on year change.

In Figure 13 the first-difference of the BP 2013 scenario to 2030 is given. This first-difference trend is markedly different to that for RCP 4.5, with a decrease shown from as soon as 2013 on.

Figure 13. Data as for Figure 11 except that the first-difference transformation of the RCP4.5 (CIMP5) model is depicted in light blue and the first-difference transformation of projected atmospheric $CO_2$ from projected anthropogenic $CO_2$ emissions (BP 2013) is added (red curve).

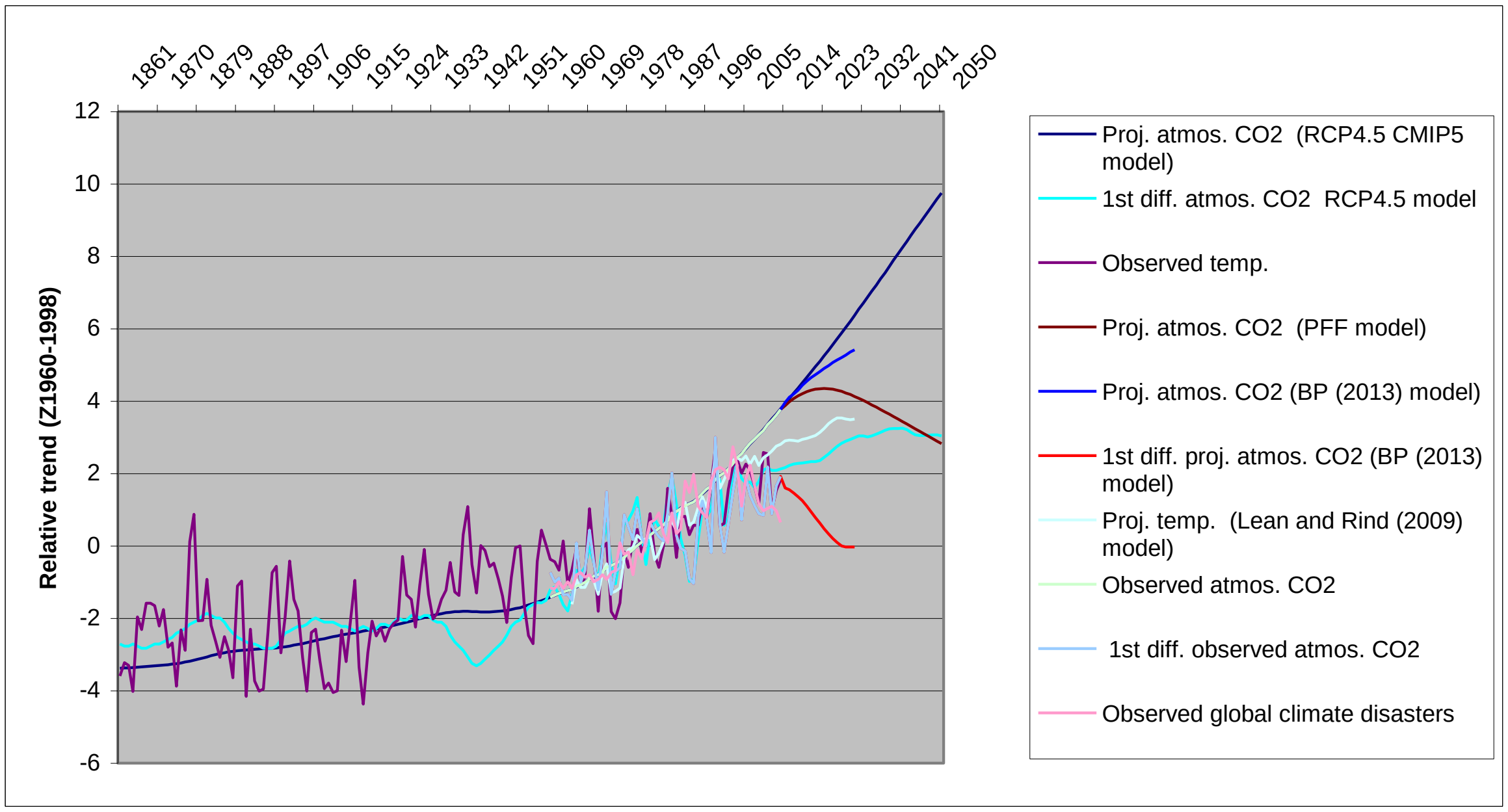

In Figure 14, the first-difference trend for the peak fossil fuel scenario to 2050 is added.

Figure 14. Data as for Figure 12 except that the first-difference transformation of projected atmospheric $CO_2$ from projected anthropogenic $CO_2$ emissions (BP 2013) is depicted in light blue and first-difference projected atmospheric $CO_2$ (peak fossil fuel model) is added (red curve).

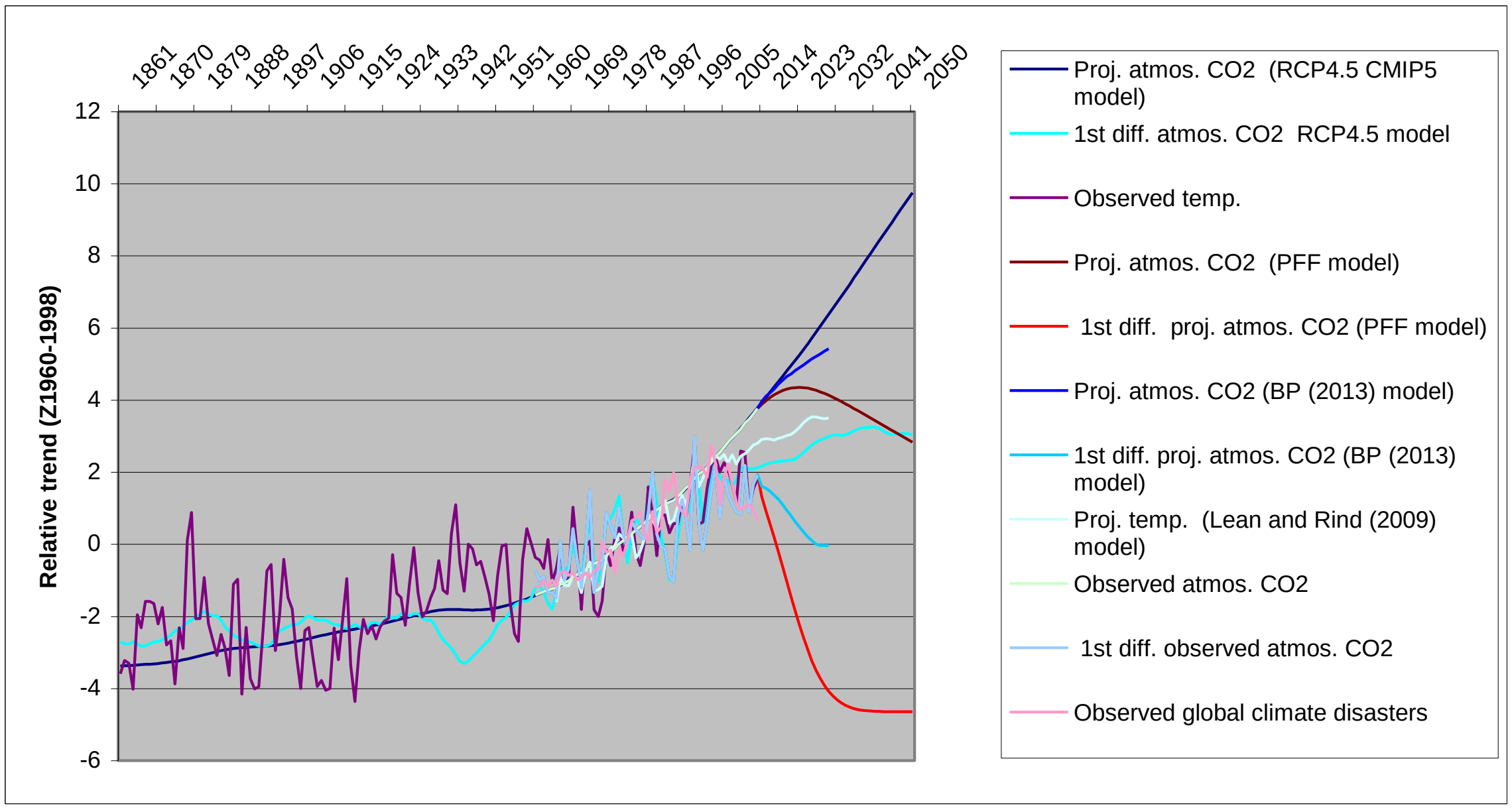

This first-difference peak fossil fuel scenario shows the most marked decrease of all the first-difference scenarios shown. The scenario seems on the face of it hard to entertain. Yet it should be borne in mind that never before over the whole period depicted from 1850 has the absolute amount of $CO_2$ in the atmosphere decreased, as happens in the peak fossil fuel scenario.

Concerning temperature, Figure 14 overall shows that it has shown considerable variation over the period from the 1850s to the present so that its change in trend since 1998 is not unprecedented. For climate disasters, however, the situation is different. Albeit over the shorter period from 1960 to the present, the decrease since 2005 is unprecedented. When the correlation between temperature and disasters and between temperature and first difference $CO_2$ from 1960-2013 is then recalled, this suggests that, taken together, the two climate categories studied - global atmospheric surface temperature and total climate disasters  - have tended to follow either the first-difference BP or the first-difference peak fossil fuel future scenarios. Entertaining this, the long time-perspective given by the figure -- commencing in 1850 - enables the observation that, the global climate is about to enter an unprecedented period.

## Discussion

In this section, the following is considered: the implications of the results for current scientific stances; precedents for use of first difference rate of change in public policy; implications of the results for scenario-based risk analysis; and implications for energy policy.

### *Implications of this work for current scientific stances*

While generating differences, a strong case can be made that this work is nonetheless consistent with current climatology.

The differences are twofold.

First, the concept that the rate of change of atmospheric $CO_2$ has a role in the trend in climate event frequency can be seen to be consistent with current climatology. This is because it is possible that the rate effect has not been revealed until now because there has not in the modern era until now been in existence a previously large monotonically increasing forcing which has started to decrease. So it would seem that climatology might *develop* because of this -- it would not be that part of current climatology was “wrong".

A second point is that the first difference view provides some very clear signature matches – statistically significant correlations – between anthropogenic inputs and climate variable outputs. This provides additional strong evidence supporting the notion of the dominant effect of human activity on current climate.

In particular, much of short term climate variation that might have been considered noise (for example, Karl et al., 2009) can now be considered signal. (While beyond the scope of this paper, the noise-like appearance of this signal opens the possibility that it might be noise from the plant biosphere photosynthesis system. In other words, noise from one system acting as a signal to another system.)

*Precedents for use of first difference rate of change in public policy*

The fact that the climate curves match the atmospheric $CO_2$ trend in first-difference form is of interest because rate-of-change curves of this type are used in public policy as leading indicators of the future performance of real world events. For example, from the OECD-published volume *OECD Composite Leading Indicators: a tool for short-term analysis:*

http://www.oecd.org/fr/std/indicateursavancesetenquetesdeconjoncture/15994428.pdf

A number of different derived measures are available in different OECD publications. These measures assist users in the analysis and interpretation of recent developments in the Composite Leading Indicator (CLI). These are…

**The 12-month percent change of the composite indicator and the 12-month trend rate for the reference series.**

The 12-month percent change of the indicator is a rate where the initial value is a 12-month centred average (the rate is calculated by dividing the figure for a given month by the 12-month moving average centred on m-12). This rate *gives early warnings of turning points in the CLI.* The timing of the peaks/troughs corresponds to the local maximum/minimum of the rate. For perfectly well-behaved cycles, this gives signals about 12 months ahead (in practice however, time series are not perfectly well-behaved)...

…

An assessment of the ability of these measures to predict turning points has been done for the United States over the period 1980-1998. Signals given by the measures have been compared to the turning points in the reference series. Different statistics have been calculated including the: number of leads/lags in months and the number of missing and extra signals… The 1-month percent change gives early signals of all turning points, except in January 1989. The 12-month percent change gives signals of all the turning points with a longer lead. Both measures give extra signals of turning points. Over the period 1980-1998, there are no missing signals.

The above excerpt shows a rate of change series being used by a well-established mainstream organisation, the OECD, to predict a future trend change in the raw series a substantial number of periods ahead.

Translating this thinking to the present topic could mean the following. If future climate event trends follow 2012 along the peak fossil fuel scenario trajectory, this will mean observed real-world climate trends will be acting as a leading indicator that peak fossil fuel is arriving within the timeframe predicted by published studies such as Nel and Cooper (2010).

Considering first-difference $CO_2$ as a leading indicator potentially enables two major insights. First, it is an answer to the question of why this hinge of history-like turn is happening now. Second, so answered, the trends are also a harbinger that peak fossil fuel is a scenario to take seriously.

*Implications of results for scenario-based risk analysis*

The use of scenario-based risk-analysis perspectives is widely supported including by the IPCC (Klein et al, 2007): "… a robust decision framework is suitable for analysing the array of future vulnerabilities to climate change." The above trends would enable a recommendation that there would be benefit from including the first-difference scenario alongside the pre-existing scenarios shown in the earlier figures in monitoring future climate outcomes and that public policy planning and implementation was such that it would be robust against the first-difference potential outcome as well as other existing potential outcomes.

It must be admitted that it is hard to believe that the proposed future trajectory from 2014 for first difference peak fossil fuel could possibly happen. But yet, in 1950, it would have been hard to believe that the rapid rise which was to come up to the year 2000 would happen, yet it did.

In the same way that some IPCC models predict an unprecedented 8 degrees hotter by 2100 , and that is not on trend to coming true, the first-difference peak fossil fuel prediction also may not come true. Nonetheless, the IPCC had to report what their model said, and so must we.

It is noted that it can be claimed that the mean IPCC model is so far not coming true (Fyfe 2013) – possibly because of unforseen feedbacks (Leggett and Ball 2014). The future may show that the same may turn out to be true for the first-difference peak fossil fuel model.

All the above said, and noting it is important to stress that the future rate of change of $CO_2$ scenario is a scenario not a prediction (not a guarantee), it can now be monitored. Forewarned is forearmed.

*Implications for energy policy*

A feature of the first differencing method is the evidence it provides that climate trends in existence now may be leading indicators to a forthcoming peak in fossil fuel production. As stated above, this possibility suggests that climate trends should be monitored against both first difference models – business as usual and peak fossil fuel – until which of the two pathways being followed becomes clear.

If future events lean to the peak fossil fuel pathway, with the implications for future climate depicted above, it is stressed that peak fossil fuel is itself a global risk. If such implications for future climate come about, it will be so because there is a new plank in the case that the peak fossil fuel risk is real.

A metaphor therefore is that you are at your doctor's. You are told that you do not have the heart disease it was thought you had. So you are happily entertaining not having to undergo the stringencies of the diet and exercise treatment which had been prescribed. But just as you do, your doctor then goes on to say that he has found you have a cancer. And the treatment for that is -- exactly the same as for the heart disease!

## Supplementary information

Two matters are dealt with in this section: (i) an investigation of the adequacy of climate disaster trend data; and (ii) the calculation of future atmospheric $CO_2$ projections from anthropogenic $CO_2$ emissions. For references, see **References** above.

### 1. Investigation of the adequacy of climate disaster trend data

Gall (2009) draws the conclusion that current global and national databases for monitoring losses from national hazards suffer from a number of limitations, which in turn lead to misinterpretation of hazard data. These biases include:

1) hazard bias, which produces an uneven representation and distribution of losses between hazard types;

2) temporal bias, which makes it difficult to compare losses across time due to less reliable loss data in past decades;

3) threshold bias, which results in an underrepresentation of minor and chronic events;

4) accounting bias, which underreports indirect, uninsured, and others losses;

5) geographic bias, which generates a spatially distorted picture of losses by over- or underrepresented certain locales; and

6)  systemic bias, which makes it difficult to compare losses between databases due to different estimation and reporting techniques.

With the above background,  the EM-DAT disaster events database (Centre for Research on the Epidemiology of Disasters, Ecole de Sante Publique, Universite Catholique de Louvain, Brussels, Belgium) (EM-DAT 2013) is chosen as it is, as mentioned, widely used, for assessment of the extent in the database of the biases listed by Gall (2009).

In the EM-DAT database, climate disaster types are coded into four main categories: extreme temperature, flood, storm, drought and wildfire. The extreme temperature category is further is made up of two main sub-types – heat wave and cold wave. In this study the heat wave sub-type is used. Trends in theses categories are assessed against the six bias risks listed by Gall (2009).

1*)  Hazard bias; 2) Temporal bias; 5) Geographic bias*

If there is temporal bias it should be less apparent in countries with on average higher GDP per capita. The following figures show relative phasings of peaks in event time series by continent and climate disaster type.

Supplementary figure 1.  Flood disasters by continent: relative trend (Z scores) and polynomial curve of best fit (total cases: 3726)

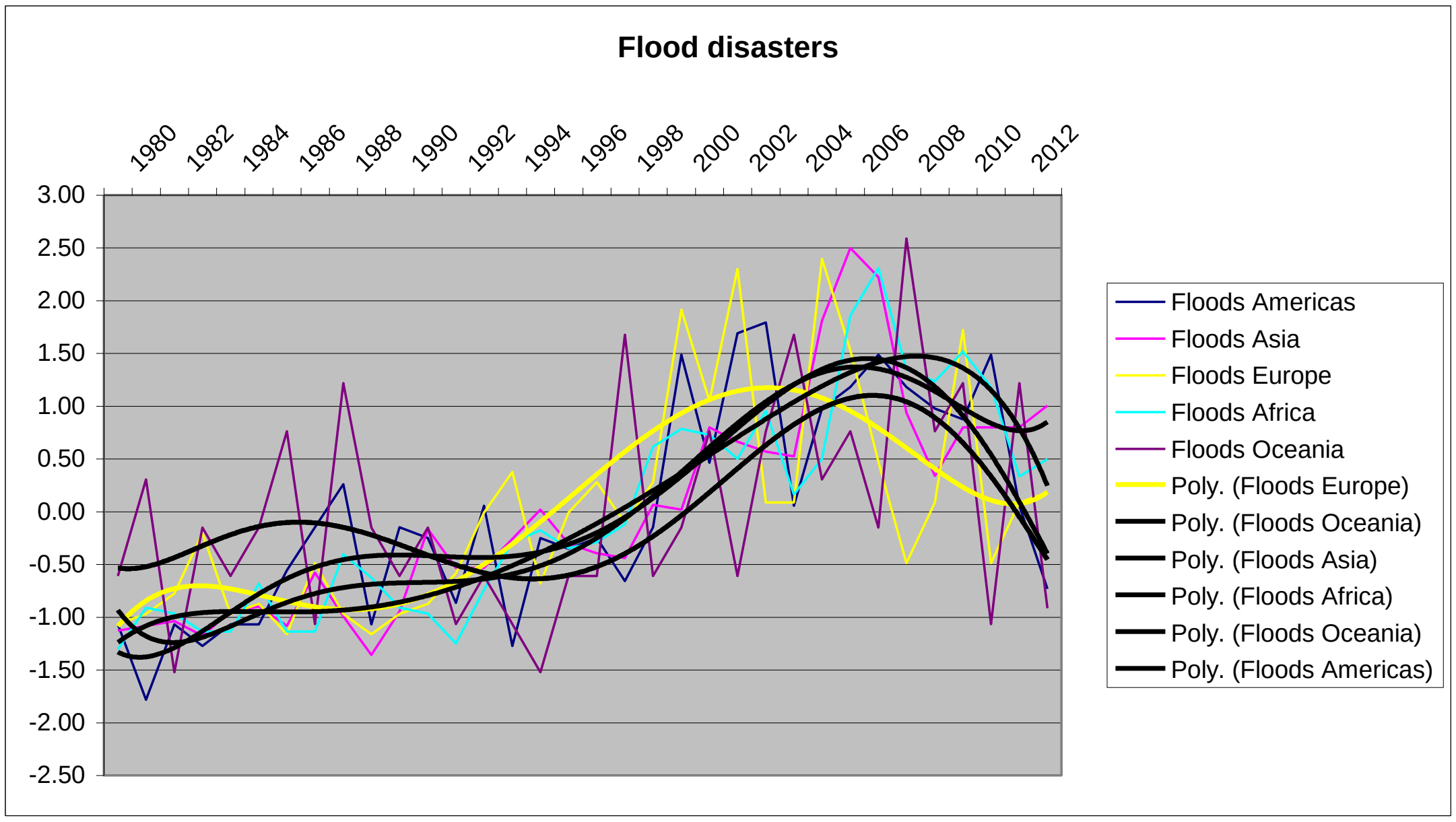

Supplementary figure 2. Storm disasters by continent: relative trend (Z scores) and polynomial curve of best fit (total cases: 2763)

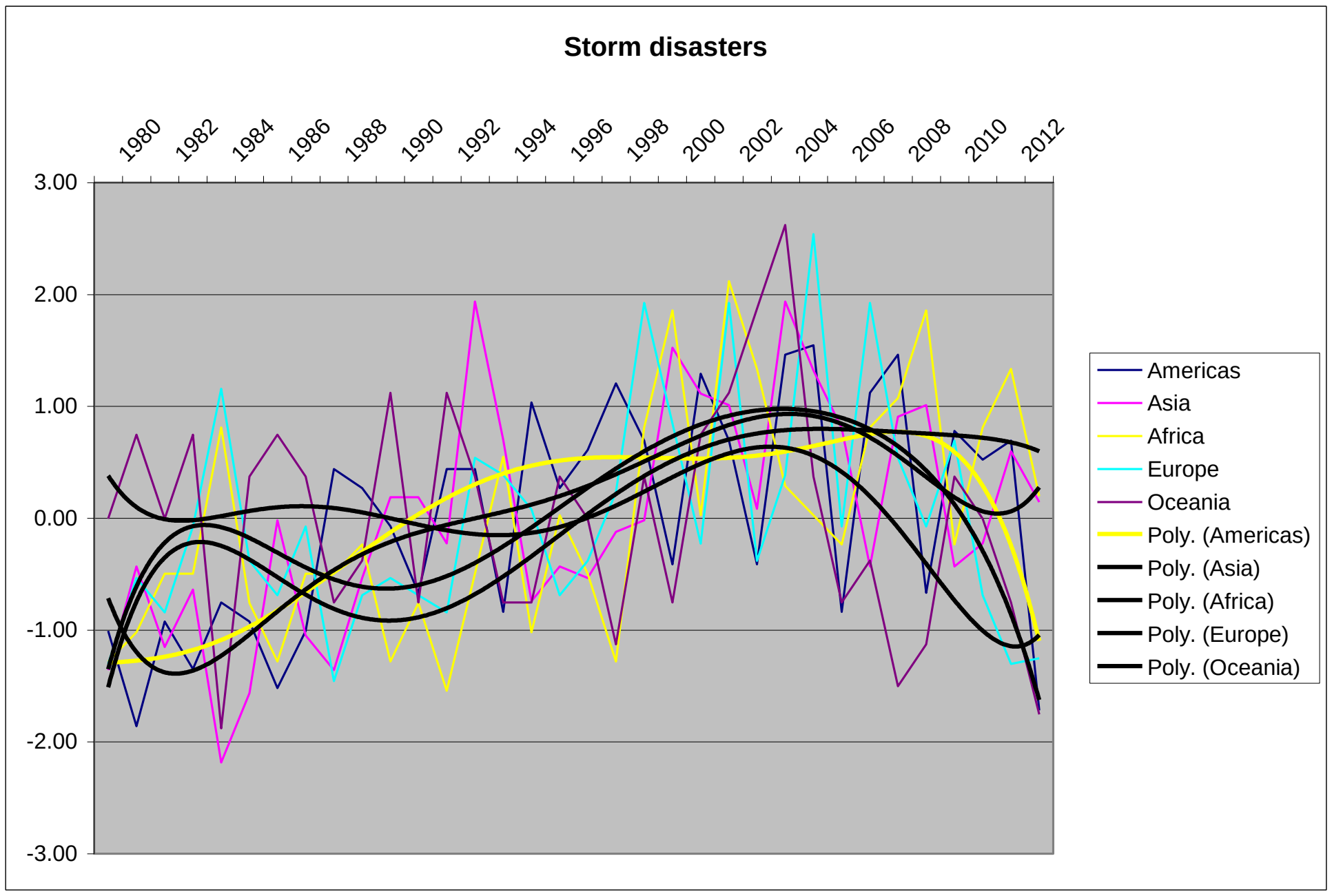

Supplementary figure 3. Drought disasters by continent: relative trend (Z scores) and polynomial curve of best fit (total cases: 449)

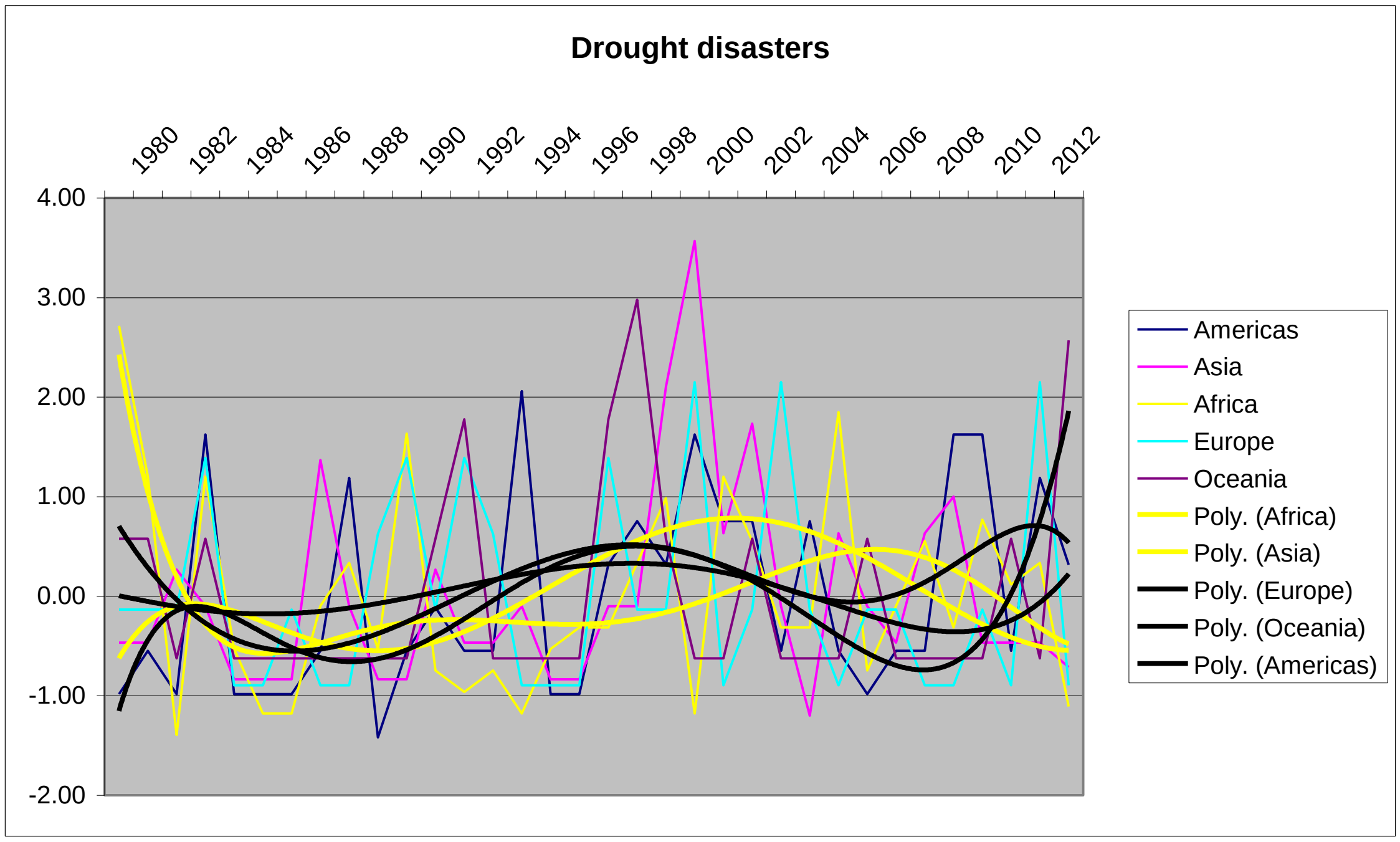

Supplementary figure 4. Wildfire disasters by continent: relative trend (Z scores) and polynomial curve of best fit (total cases: 332)

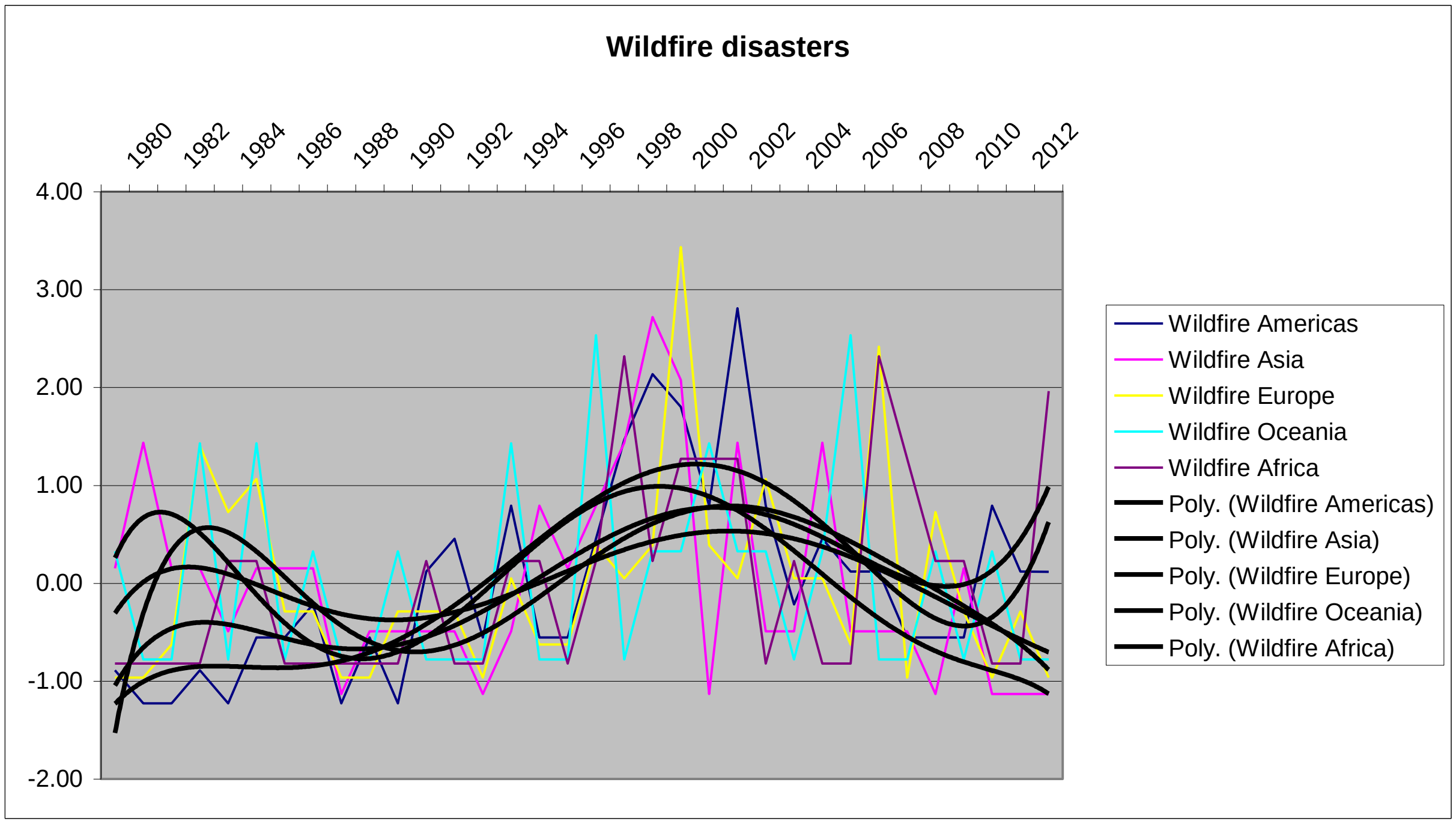

Supplementary figure 5. Heatwave disasters by continent: relative trend (Z scores) and polynomial curve of best fit (total cases: 140)

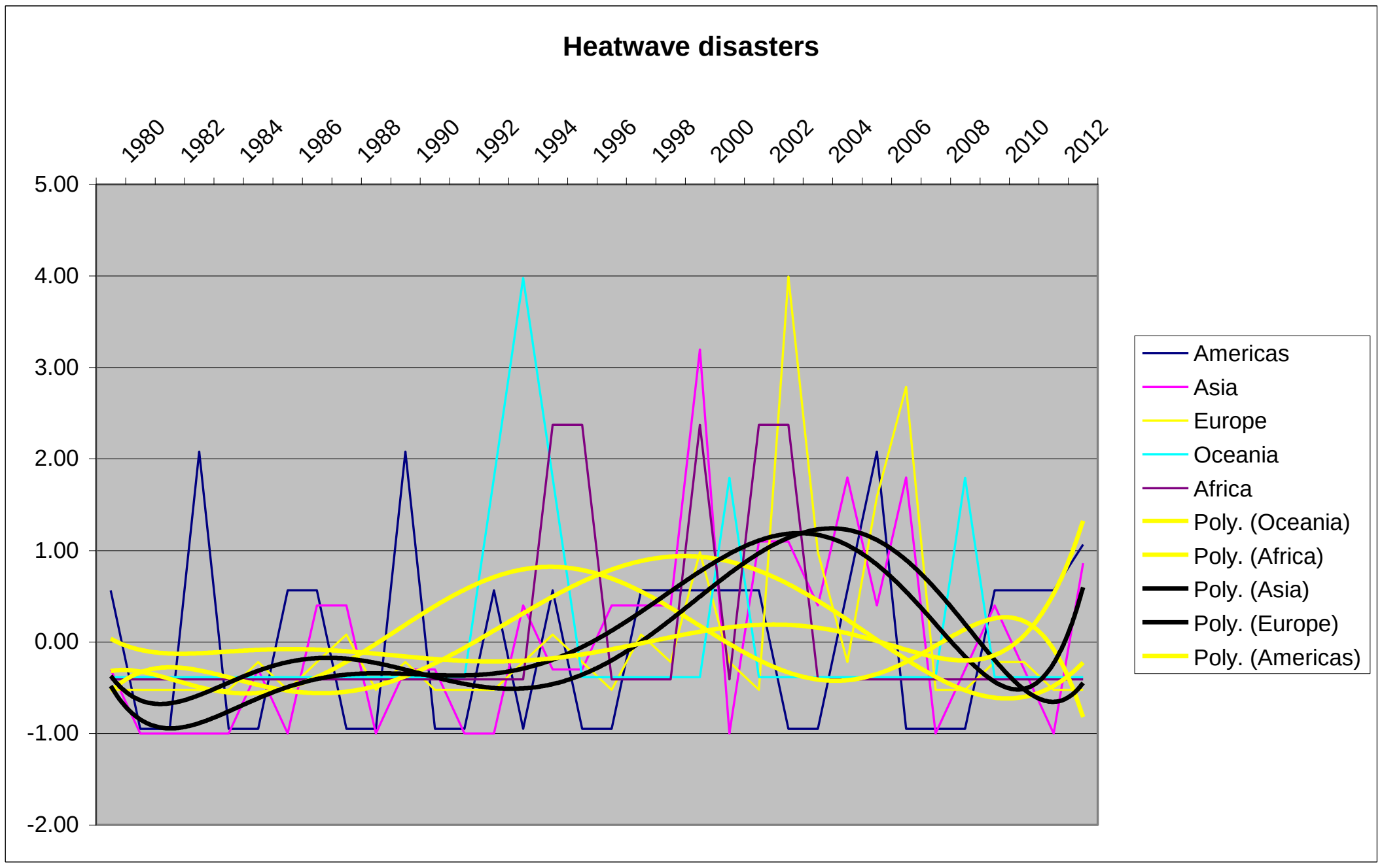

The preceding figures show that climate disaster categories tend to show single peak years. These peak years tend to differ by disaster type, but per disaster type, with a minority of exceptions, to coincide across continents. This tendency is greater for disasters for which there is the largest number of cases. These results are summarised in Supplementary table 1.

Supplementary table 1. Disaster types: total reported and peak years by continent: per cent aligned

| | **No. reported disasters globally 1980-2013** | **Peak years by continent: per cent aligned** |
|---|---|---|
| | | |
| Flood | 3726 | 80 |
| Storm | 2763 | 80 |
| Drought | 499 | 60 |
| Wildfire | 332 | 100 |
| Heat wave | 140 | 40 |

*2) Temporal bias*

Barredo (2009) noted that a simple assessment of the number of damaging events included in the flood data he studied built up from the EM-DAT database and other sources revealed that in the first half of the assessment (1970–1988) there were 32 events, whereas in the period 1989–2006 there were 90. He suggested that this difference went reasonably beyond natural variability or societal changes and could therefore be attributed to inaccuracies in the accounting of the events.

This question is explored further by comparing the disaster event trend with that of an external trend.

In doing so, we note that it is considered that records are reasonably complete for major disaster types in some countries from 1970, so similarity with a valid external comparator should occur at least from this start date.

It can be seen that in Supplementary figure 6 covering the period 1960 to 2003 climate disasters show a relationship to the level of $CO_2$ which is very close to linear. This supports the above statement of reliable data from 1970, and further is support for taking that view back at least to 1960. Hence this start year is used in this study.

**Supplementary figure 6**

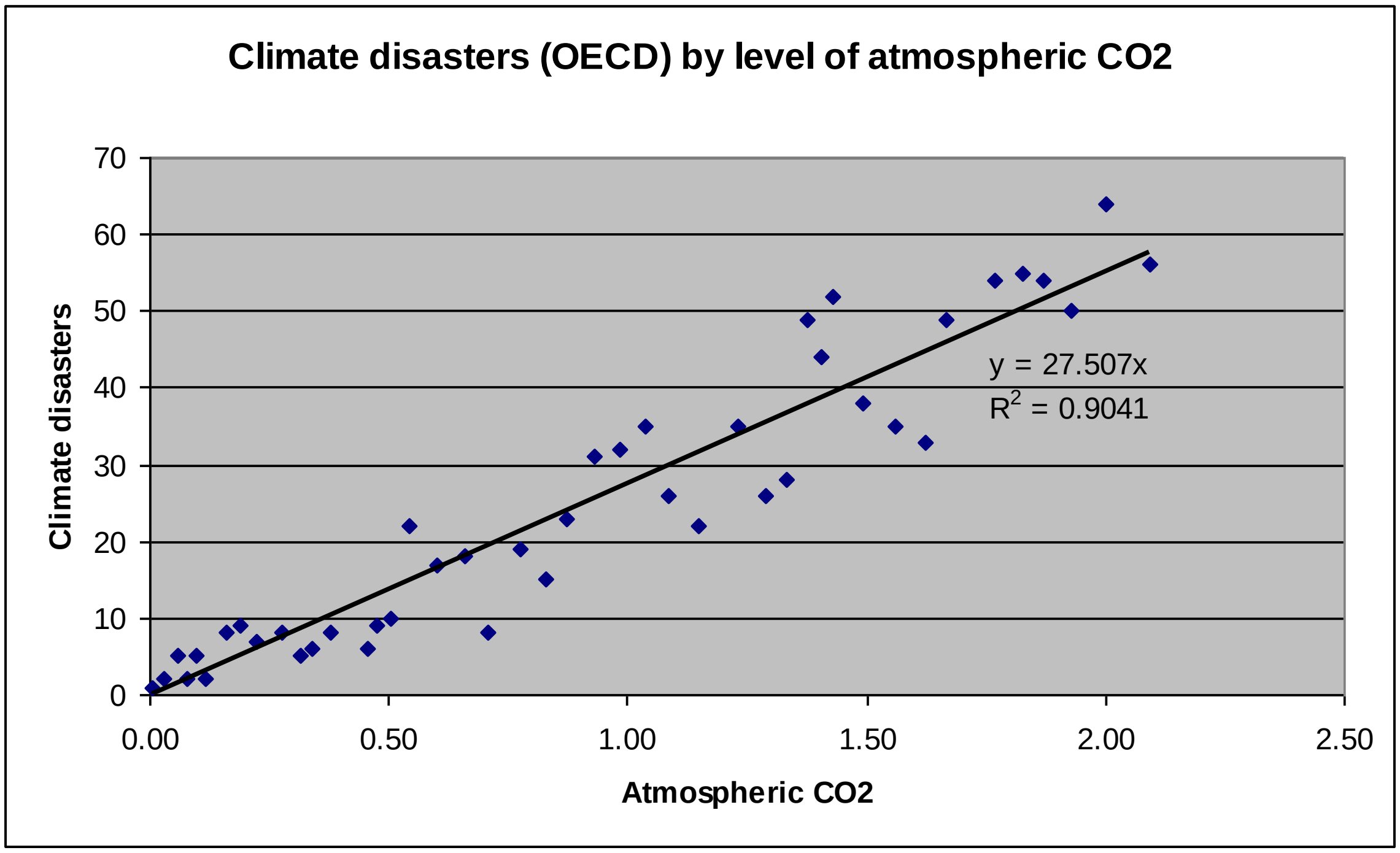

How does data from the rest of the world compare? Supplementary figure 7 shows that both curves increase monotonically to the early 2000s, and as expected the non-OECD curve is somewhat more non-linear that the OECD curve. Nonetheless both curves show a decrease as the 2000s wear on (see main account).

Supplementary figure 7: Number of climate disasters 1960-2013 by country grouping

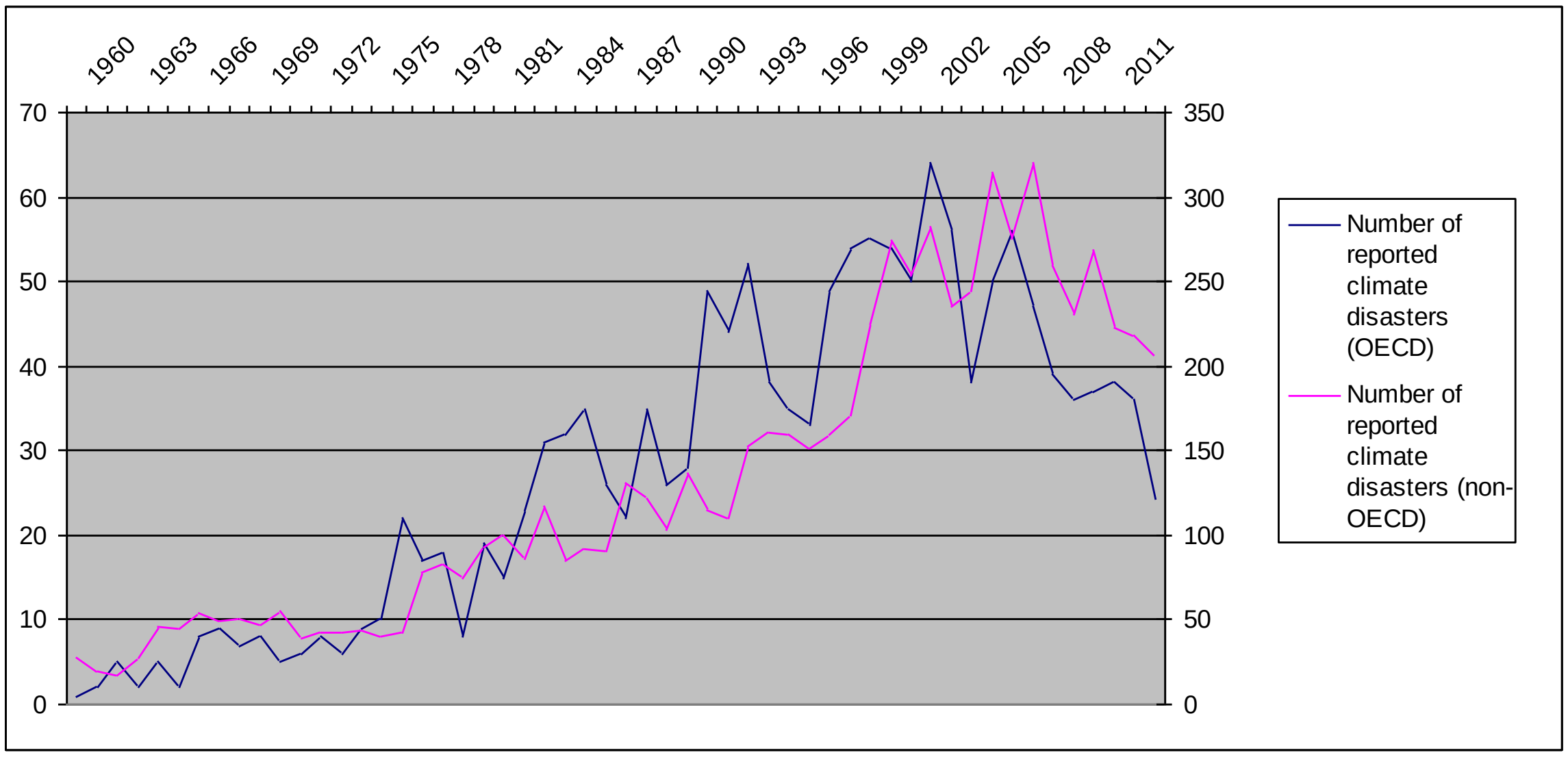

*3) Threshold bias; 4) accounting bias*

EM-DAT focuses on major disasters. These are more likely to show up in records, hence reducing threshold and accounting bias.

*Conclusion to section on disaster data*

From the positive results from the above eight tests, it is considered that the OECD climate disaster data from the EM-DAT database is fully adequate for use as an extra climate outcome alongside global surface temperature in the assessment of climate trends and their relationship to atmospheric $CO_2$. In the assessment, as for the preceding tests, the number of events per year for each of the categories of flood, storm, drought, wildfire and heat wave is added to produce an annual aggregate climate disaster time series.

**2. Calculation of future atmospheric $CO_2$ projections from anthropogenic $CO_2$ emissions**

Calculation of future atmospheric CO2 projections from anthropogenic CO2 emissions is done by simple regression from the prior relationship between anthropogenic $CO_2$ emissions and atmospheric $CO_2$. Supplementary figure 8 shows the data used. It can be seen that for the different estimates available there is a broadly linear relationship over the period of overlap.

Supplementary figure 8: Trends observed and projected for global anthropogenic CO2 emissions and atmospheric CO2

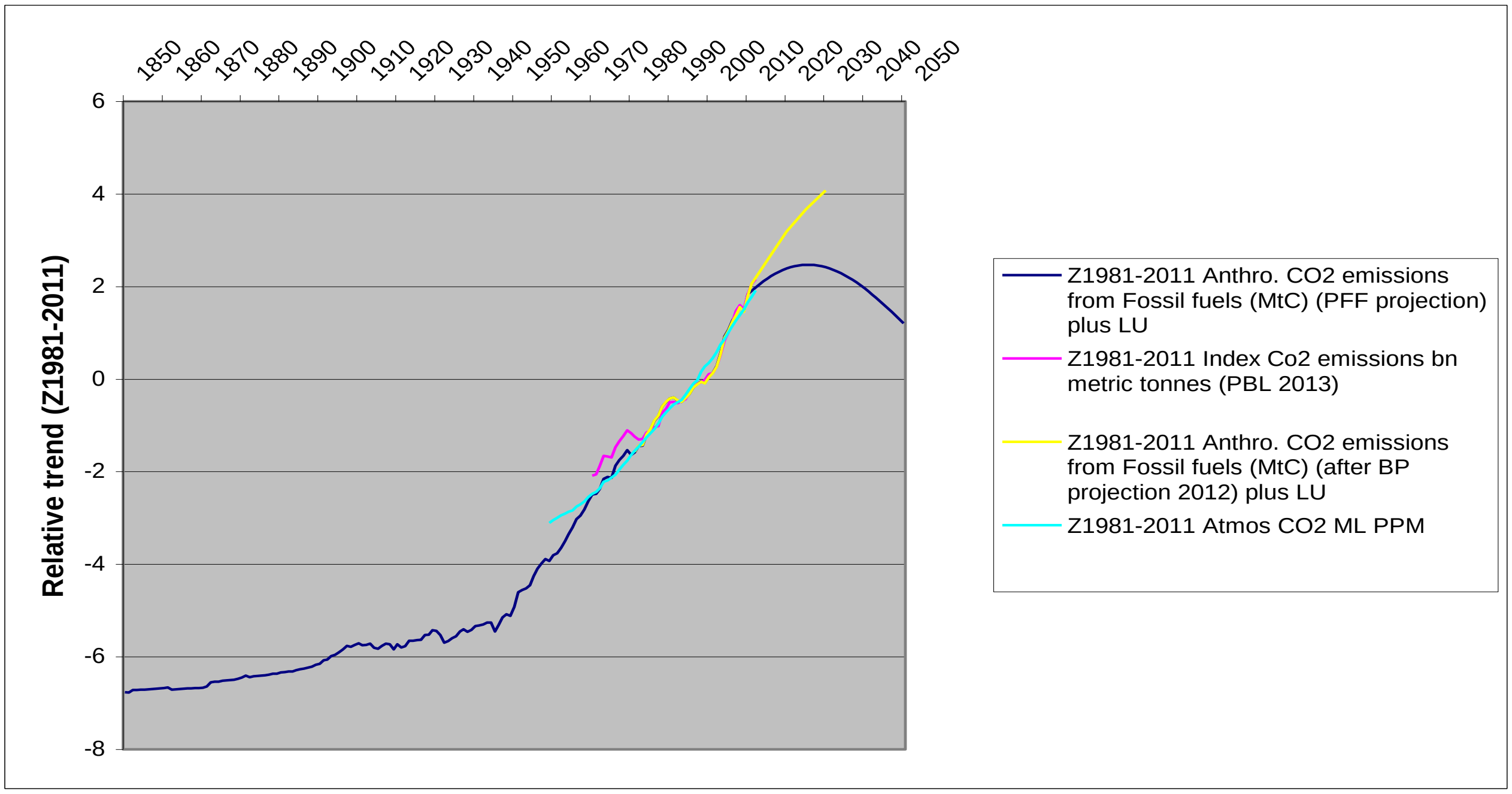

Supplementary figure 9 shows a scatter plot of the relationship between anthropogenic $CO_2$ emissions and atmospheric $CO_2$ for the period 1959 to 2012. The correlation is large and the statistical significance high (R squared = 0.989; P = 1.7E-44).

Supplementary figure 10 shows the predicted future atmospheric CO2 based on the relationship shown in Supplementary figure 9. This is the data used in the analysis in the paper proper.

Supplementary figure 9. Scatter plot of the relationship between anthropogenic $CO_2$ emissions and atmospheric $CO_2$ for the period 1959 to 2012

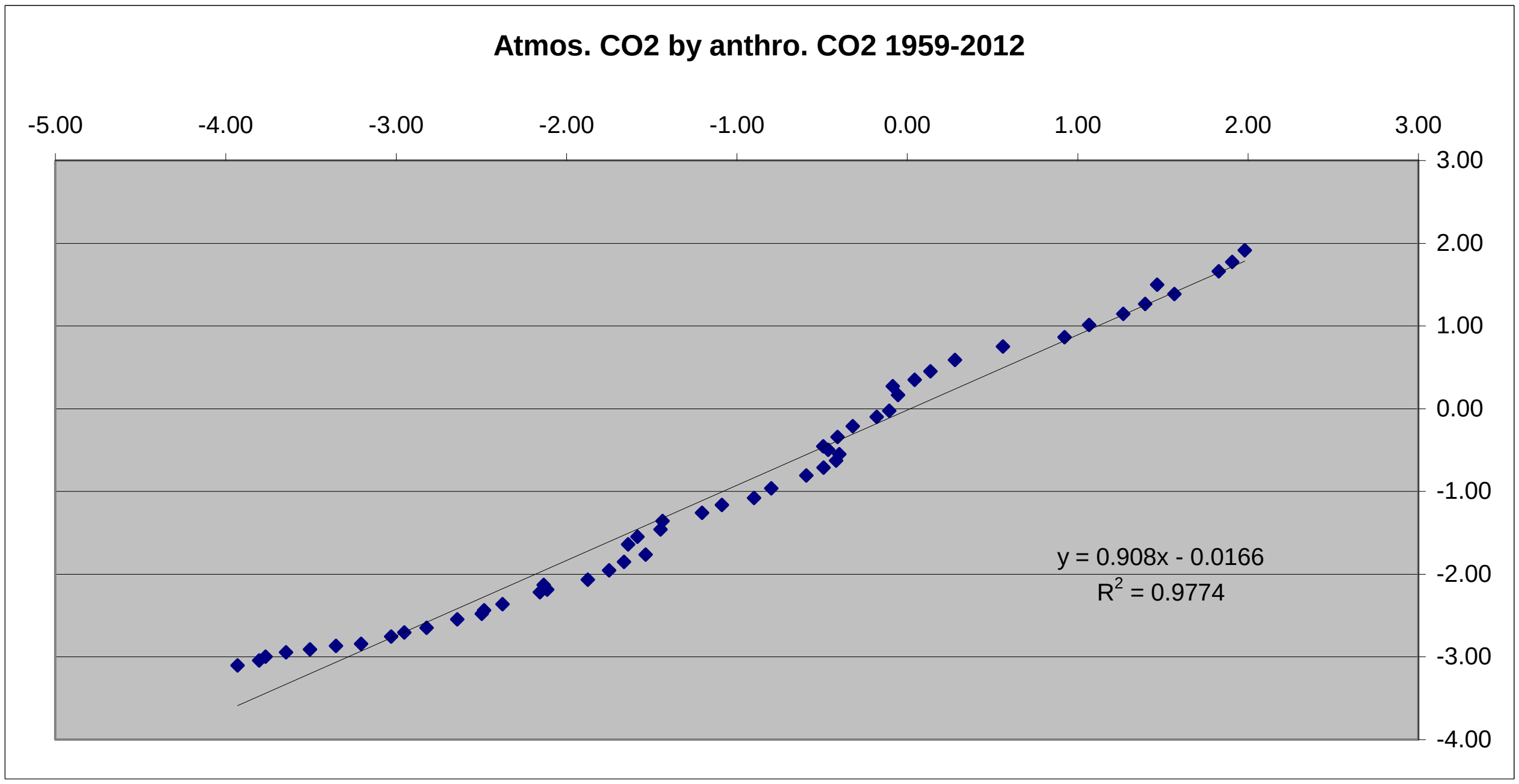

Supplementary figure 10. Predicted future atmospheric CO2 based on the relationship shown in Supplementary figure 9.

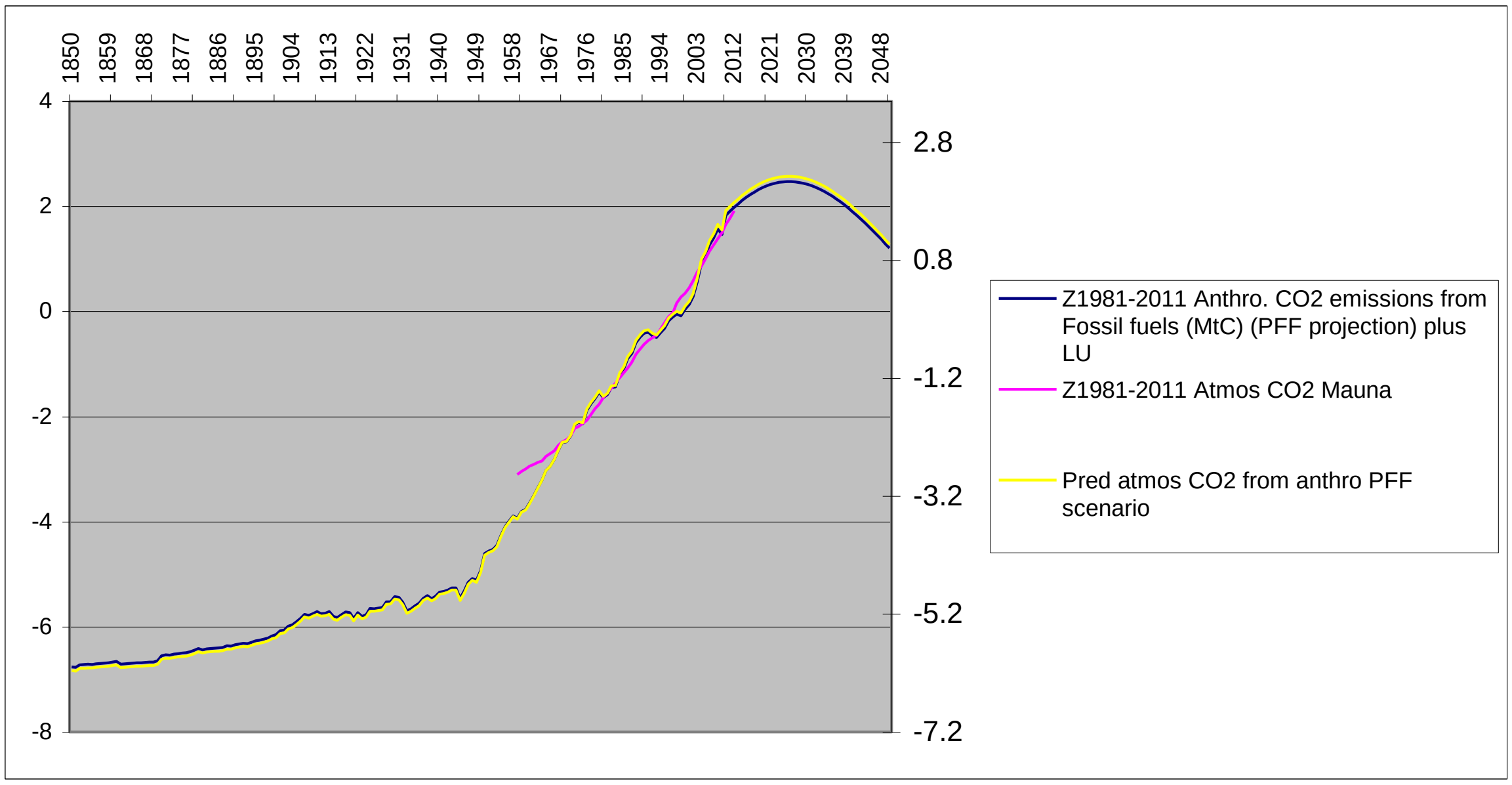

## Supplementary references

For references, see **References** above.